\documentclass{article}

\usepackage{amsmath}
\usepackage{amsfonts}
\usepackage{amssymb}
\usepackage{amsthm}
\usepackage{latexsym}
\usepackage{ifthen}
\usepackage{graphicx}
\usepackage{url}

\newtheorem{proposition}{Proposition}%
\newtheorem{lemma}[proposition]{Lemma}
\newtheorem{theorem}[proposition]{Theorem}
\newtheorem{corollary}[proposition]{Corollary}

\newcommand{\supp}{\text{supp}}
\newcommand{\cat}{\text{CAT}}

\title{The space of equidistant phylogenetic cactuses}

\date{\today}

\author{Katharina T. Huber\\University of East Anglia, School of Computing Sciences\\Norwich, NR4 7TJ, UK\\k.huber@uea.ac.uk
\and Vincent Moulton\\University of East Anglia, School of Computing Sciences\\ Norwich, NR4 7TJ, UK\\v.moulton@uea.ac.uk
\and Megan Owen\\Lehman College, CUNY, Department of Mathematics\\New York, NY 10468, US\\megan.owen@lehman.cuny.edu
\and Andreas Spillner\\Merseburg University of Applied Sciences\\ 06217 Merseburg, Germany\\andreas.spillner@hs-merseburg.de
\and Katherine St. John\\Hunter College, CUNY, Department of Computer Science\\New York, NY 10065, US\\katherine.stjohn@hunter.cuny.edu}

\begin{document}

\maketitle

\begin{abstract}
We introduce and investigate the space of 
\emph{equidistant} $X$-\emph{cactuses}. These are 
rooted, arc weighted, phylogenetic networks with leaf set $X$, 
where~$X$ is a finite set of species, and all leaves have the same
distance from the root. The space contains as a subset the space of 
ultrametric trees on $X$ that was introduced by Gavryushkin and Drummond.
We show that equidistant-cactus space is a \cat(0)-metric space which
implies, for example, that there are unique geodesic paths between points.
As a key step to proving this, we present a combinatorial result
concerning \emph{ranked} rooted $X$-cactuses. In particular, we show that 
such networks can be encoded in terms of a pairwise compatibility
condition arising from a poset of collections of 
pairs of subsets of~\(X\) that satisfy certain set-theoretic properties.
As a corollary, we also obtain an encoding of ranked, rooted
$X$-trees in terms of partitions of~$X$, which provides
an alternative proof that the space of 
ultrametric trees on $X$ is~\cat(0).
As with spaces of phylogenetic trees, we expect that our results
should provide the basis for and new directions in performing statistical 
analyses for collections of phylogenetic networks with arc lengths.
\end{abstract}


\section{Introduction}
\label{sec:introduction}

Currently, there is great interest in developing theory and techniques
to understand and construct (rooted) \emph{phylogenetic networks}.
Generally speaking, for a set of species, such a network 
consists of a rooted, directed acyclic  graph and a bijective map 
from the species to the set of sinks of the graph (in case the
graph is a tree, the network is called a (rooted) \emph{phylogenetic tree}). 
Phylogenetic networks are important as they
can be used to represent the evolutionary history of species
that cross with one another (through evolutionary processes 
such as hybridization and recombination). To date, much of the research on 
phylogenetic networks has focused on understanding the
structure of special types of networks and ways to 
build them (see \cite{S16a} for a recent overview of the area).
More recently, however, as the theory for phylogenetic networks has developed,
there has been growing interest in understanding how to 
equip collections of phylogenetic networks with suitable metrics, 
giving rise to so-called \emph{network spaces}. 
As has been demonstrated for the intensively studied spaces of phylogenetic 
trees (cf. e.g. \cite{BHV01,GD16}, and the review \cite{st2017shape}),
or \emph{tree-spaces}, this 
point of view is valuable as it provides insights 
into statistical approaches to analyze and systematically compare networks.

Network spaces essentially come in two types: discrete
and continuous. In \emph{discrete} spaces, the elements
of the space are distinct, non-isomorphic networks, and 
a metric is commonly given by defining
the distance between two networks to be the length
of a minimal sequence of local network operations that converts
one network into the other. In \emph{continuous} spaces, the arcs in the
networks have non-negative, real-valued lengths and one
network can be converted into
the other by shrinking or lengthening arcs in a continuous manner.
To date, nearly all results on network spaces have 
concerned discrete spaces (see, for example, 
\cite{BSC17a,GVJ17a,JJEIS18a},
for related results on discrete spaces 
of unrooted networks see e.g. \cite{HMW16a}).  Indeed, to 
the best of our knowledge, very few results have been
presented on continuous network
spaces except for the recently introduced spaces of
(unrooted) circular split networks\footnote{Strictly 
speaking, these spaces should probably be thought of
as ``spaces of circular split collections''.} \cite{DP17a}.
This is probably in part because the study of phylogenetic
networks with arc lengths is somewhat less developed than
the study of those without.

\begin{figure}
\centering
\includegraphics[scale=1.0]{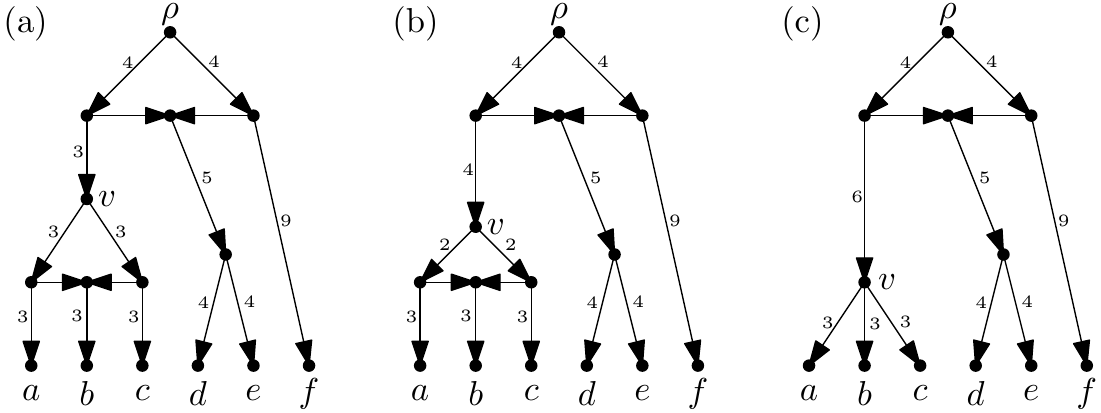}
\caption{(a) An \(X\)-cactus for \(X = \{a,b,c,d,e,f\}\)
with root~\(\rho\) that is equidistant since every
directed path from~\(\rho\) to a sink has the same length, namely~\(13\).
All incoming arcs at vertices with indegree~2 have length~0 and are
drawn horizontally.
(b) The rooted \(X\)-cactus obtained by lengthening the incoming arc
and shrinking the outgoing arcs at vertex~\(v\) by~1.
(c) The rooted \(X\)-cactus obtained by continuing the lengthening
and shrinking of the arcs at vertex~\(v\) until both outgoing arcs
have length~0, contracting the cycle below~\(v\) completely.}
\label{fig:example:x:cactus:introduction}
\end{figure}

In this paper, we introduce a new continuous space of
phylogenetic networks that can be regarded as a generalization 
of the \(\tau\)-\emph{space} of ultrametric trees
that was introduced in \cite{GD16}.
For a set~$X$ of species, our network space $\mathfrak{N}(X)$
is comprised of \emph{equidistant} \(X\)-\emph{cactuses} 
(see Figure~\ref{fig:example:x:cactus:introduction}(a)
for an example of such a network).
A rooted $X$-cactus is essentially a rooted phylogenetic network in which 
no two distinct cycles in the underlying graph have an arc in common.
Note that if all vertices of a rooted \(X\)-cactus have indegree at most~1 the 
network is just a \emph{rooted phylogenetic} \(X\)-\emph{tree}.
The extensively studied class of (rooted) \emph{level-1 networks}
(see e.g. \cite{RV09a}) also provides examples of rooted $X$-cactuses. 
Assigning a non-negative real-valued length to each of 
the arcs in a rooted phylogenetic network,
then such a network~\(\mathcal{N}\) is called \emph{equidistant} if, 
for any fixed vertex~\(v\) of~\(\mathcal{N}\),
all directed paths from~\(v\) to any sink of~\(\mathcal{N}\)
have the same length. Algorithms for constructing
equidistant phylogenetic networks 
have been studied in, e.g.,~\cite{BT16a} and \cite{CJLY06a}. 

Following one of the common approaches used to construct 
tree-spaces, we define equidistant-cactus space $\mathfrak{N}(X)$ 
in terms of an \emph{orthant space} (see e.g.~\cite{MOP15a}).
Basically, an orthant space is a collection of real
orthants that are glued together
along their boundaries and
that is equipped with the metric induced by using the
Euclidean metric within each orthant. That is, the distance
between two points in the same orthant is the Euclidean distance
between these points, and the distance between two points in different
orthants is the length of a shortest path, or \emph{geodesic path},
between these points. The length of such a path is computed by summing
the Euclidean lengths of the restrictions of the path to each orthant.
In particular, each pair of points in~$\mathfrak{N}(X)$ represents two 
equidistant \(X\)-cactuses, and moving along a geodesic path
between the points continuously converts
one \(X\)-cactus into the other by shrinking and lengthening
arcs (see Figure~\ref{fig:example:x:cactus:introduction}(b) and~(c)),
which may also result in a change of the length of the paths from the
root to the sinks.
Note that the points of \(\tau\)-space
correspond bijectively to equidistant $X$-trees and that it can be constructed by
gluing together orthants indexed by \emph{ranked phylogenetic trees}.
We take a similar approach to define $\mathfrak{N}(X)$, indexing orthants instead by 
\emph{ranked} \(X\)-\emph{cactuses}, in which a ranking of the
vertices that respects the direction of the arcs in the rooted \(X\)-cactus is given.
We remark that ranked phylogenetic networks have been
recently introduced and that research has focused on
counting and enumerating certain classes
of such networks (see e.g. \cite{BLS20a,CFY21a} and the references therein).

A critical aspect that influenced our construction of $\mathfrak{N}(X)$ 
was that -- as has been shown for \(\tau\)-space \cite{GD16} -- we wanted it to be a \emph{\cat(0)-metric space}.
Being \cat(0) is an important geometrical property that has been exploited in various 
applications within phylogenetics and beyond  (see e.g. \cite{AM20a}).
A space being \cat(0) immediately implies that 
there is a \emph{unique} geodesic path between any two points, a property
that underpins many useful computations that can be performed for tree- and orthant-spaces.
More specifically, approximations of the median as well as of the Fr\'echet mean
and variance can be computed in complete \cat(0)-metric spaces,
which include \cat(0)-orthant spaces \cite{MOP15a,bacak2014computing};
a central limit theorem holds for \cat(0)-orthant spaces \cite{barden2018logarithm};
and methods for computing confidence sets \cite{willis2019confidence}
and an analogue of partial
principal component analysis \cite{nye2017principal,nye2014algorithm}
can be directly extended from 
the unrooted tree space presented in \cite{BHV01}
to \cat(0)-orthant spaces.
Most of this paper is devoted to proving a crucial combinatorial 
result concerning rooted $X$-cactuses (Theorem~\ref{theo:chains:networks:1:to:1}) which implies, via a classical result of Gromov
for orthant spaces, that $\mathfrak{N}(X)$ is~\cat(0).
In passing, we remark that the space of networks described in~\cite{DP17a}
is not a \cat(0)-metric space.

The rest of this paper is structured as follows.
In Section~\ref{sec:preliminaries}, we formally define rooted \(X\)-cactuses as
well as some related concepts. In Section~\ref{sub:sec:rankings}, 
we then introduce rankings of rooted \(X\)-cactuses and 
equidistant $X$-cactuses, which are both defined in terms of so-called
time-stamp functions.
As well as characterizing when a 
rooted \(X\)-cactus admits a ranking of its vertices that is 
consistent with the direction of its arcs, we make an important 
observation concerning ranked $X$-cactuses (Lemma~\ref{lem:size:ranking}), 
which implies that the maximal chains in a certain poset mentioned
in the next paragraph
all have the same length, i.\,e.\, $\lvert X \rvert-1$.
In Section~\ref{sec:equidistant:x:trees}, we use the simpler
case of equidistant \(X\)-trees to outline our approach
for the construction of a network space that is~\cat(0),
including a new proof that \(\tau\)-space is~\cat(0). 

In Section~\ref{sec:encoding}, we describe how 
ranked \(X\)-cactuses give rise to \emph{set pair systems}
as defined in \cite{HMS21a}
and present the properties that characterize set
pair systems that arise from ranked \(X\)-cactuses.
We also define a binary relation on general set pair systems,
and, in Section~\ref{sec:poset:p:x}, we establish that 
this relation yields a bounded graded poset on the
set pair systems that arise from ranked \(X\)-cactuses. 
In Section~\ref{sec:encoding:ranked:x:network},
we establish our main combinatorial result
(Theorem~\ref{theo:chains:networks:1:to:1}), namely that 
chains in this poset encode ranked \(X\)-cactuses. In simpler terms, this
can be regarded as a ``pairwise compatibility"  result for set pair systems,
which is analogous to the well-known \emph{Splits Equivalence Theorem} 
for unrooted phylogenetic trees (see e.g. \cite[Theorem 3.1.4]{SS03a}).
Using our encoding for ranked \(X\)-cactuses, 
in Section~\ref{sec:network:space} we construct the space $\mathfrak{N}(X)$
of equidistant $X$-cactuses and show that it is a \cat(0)-metric space.
We conclude in Section~\ref{sec:conclusion} 
by mentioning some directions for future work.

\section{Preliminaries}
\label{sec:preliminaries}

In this section, we define rooted $X$-cactuses and some related concepts
that we use later. 
We begin by recalling some standard concepts from graph theory.
A \emph{directed graph} \(N=(V,A)\) consists of a finite non-empty set
\(V\) and a subset \(A \subseteq V \times V\). The elements of
\(V\) and \(A\) are referred to as \emph{vertices} and \emph{arcs}
of~\(N\), respectively. A directed graph~\(N\) is \emph{acyclic} if there is
no directed cycle in~\(N\). Moreover, a directed acyclic graph (DAG)
\(N\)~is \emph{rooted} if there exists
a vertex \(\rho \in V\) with indegree~\(0\), called the \emph{root} of \(N\),
such that for every \(u \in V\) there is a directed path from \(\rho\) to \(u\).
In a rooted DAG, a \emph{leaf} is a vertex with outdegree~0,
an \emph{internal vertex} is a vertex with outdegree at least~1,
a \emph{tree vertex} is a vertex with indegree at most~1 and
a \emph{reticulation vertex} is a vertex with indegree at least~2.
Note that, by definition, the root of a rooted DAG is a tree vertex.
Moreover, in a rooted DAG~\(N\), we call a vertex \(v\) a \emph{child}
of a vertex \(u\) and, similarly, \(u\) a \emph{parent} of \(v\)
if \((u,v)\) is an arc of~\(N\). The set of children of a vertex~\(u\)
is denoted by \(ch(u)\).
A \emph{reticulation cycle} \(\{P,P'\}\) 
in a rooted DAG consists of two distinct directed paths \(P\) and \(P'\)
such that \(P\) and \(P'\) have the same start vertex and the same end vertex
but no other vertices in common.

\begin{figure}
\centering
\includegraphics[scale=0.9]{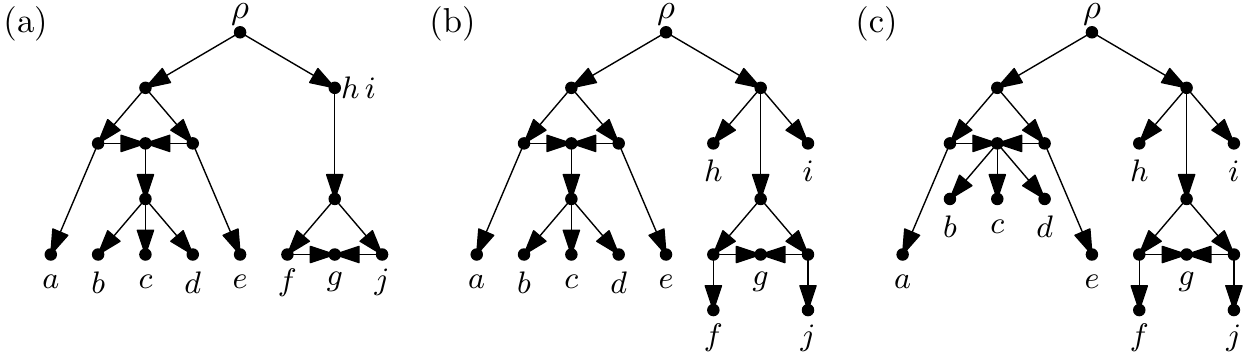}
\caption{(a) A rooted \(X\)-cactus~\(\mathcal{N}\) for \(X=\{a,b,c,\dots,j\}\)
(b) The rooted, phylogenetic \(X\)-cactus~\(\widehat{\mathcal{N}}\).
(c) The rooted, compressed, phylogenetic \(X\)-cactus~\(\widehat{\mathcal{N}}^*\).}
\label{fig:one:nested:x:networks}
\end{figure}

Let \(X\) be a finite non-empty set. A \emph{rooted} \(X\)-\emph{cactus}
\(\mathcal{N} = (N,\varphi)\)
is a rooted DAG \(N=(V,A)\) together with a map
\(\varphi: X \rightarrow V\) such that
\begin{itemize}
\setlength{\itemindent}{15pt}
\item[(RC1)]
all vertices of~\(N\) have indegree at most~2,
\item[(RC2)]
no two distinct reticulation cycles in \(N\) have an arc in common, and
\item[(RC3)]
the image \(\varphi(X)\) contains all leaves and all
tree vertices of~\(N\) with outdegree~\(1\) of~\(N\).
\end{itemize}
In Figure~\ref{fig:one:nested:x:networks}(a) we give an example of
a rooted \(X\)-cactus. We remark that if \(\lvert X \rvert = 1\)
a rooted \(X\)-cactus consists of a single vertex only.  
For better readability, we will often refer to the
vertices and arcs of~\(N\) as the vertices and arcs of~\(\mathcal{N}\).
A rooted \(X\)-cactus~\(\mathcal{N}\)
is \emph{phylogenetic}\footnote{A phylogenetic \(X\)-cactus 
is also known as a 
rooted 2-hybrid, 1-nested phylogenetic network~\cite{RV09a},
but for simplicity
we prefer to call it a rooted \(X\)-cactus since
if the root and directions are ignored we obtain an unrooted $X$-cactus \cite{HHMM20a}.} 
if \(\varphi\) is a bijection
between~\(X\) and the set of leaves of~\(\mathcal{N}\).
Note that a rooted phylogenetic \(X\)-cactus may contain leaves that are
reticulation vertices. 
A rooted \(X\)-cactus is \emph{binary} if it is phylogenetic,
all leaves of~\(\mathcal{N}\)
are tree vertices, the root has outdegree~2 and every other
internal vertex has either indegree~1 and outdegree~2 or
indegree~2 and outdegree~1.
A rooted \(X\)-cactus~\(\mathcal{N}\) is \emph{compressed} if
\(\varphi(X)\) also contains all reticulation vertices with outdegree~1
(see~\cite[p.~251]{S16a} for the concept of compression
in more general phylogenetic networks).
Rooted, compressed, phylogenetic \(X\)-cactuses as defined here correspond
to 1-nested phylogenetic networks as defined in~\cite{HMS21a}.
Note that a rooted, binary \(X\)-cactus that contains at least one reticulation vertex
cannot be compressed. A rooted \(X\)-cactus without any reticulation vertices
is called a \emph{rooted} \(X\)-\emph{tree}. Note that rooted \(X\)-trees as
defined here are in one-to-one correspondence
with the rooted \(X\)-trees as defined in~\cite{SS03a} where
the root is required to have outdegree~1.

In Section~\ref{sec:encoding:ranked:x:network},
we will need to associate with every rooted \(X\)-cactus 
\(\mathcal{N} = ((V,A),\varphi)\) a rooted, phylogenetic \(X\)-cactus
\(\widehat{\mathcal{N}} = ((\widehat{V},\widehat{A}),\widehat{\varphi})\)
as follows: For every \(x \in X\) such that \(\varphi(x)\) is not a leaf of
\(\mathcal{N}\) or such that there exists some \(y \in X \setminus \{x\}\)
with \(\varphi(y) = \varphi(x)\)
we add a new vertex~\(u\) to~\(V\), add the arc
\((\varphi(x),u)\) to \(A\), and put \(\widehat{\varphi}(x) = u\).
For all other \(x \in X\) we put \(\widehat{\varphi}(x) = \varphi(x)\).
The resulting set of vertices and arcs, respectively,
are denoted by \(\widehat{V}\) and \(\widehat{A}\) 
(see Figure~\ref{fig:one:nested:x:networks}(b)).
In addition, we associate with the resulting rooted, phylogenetic
\(X\)-cactus \(\widehat{\mathcal{N}}\) the
rooted, compressed, phylogenetic \(X\)-cactus
\(\widehat{\mathcal{N}}^* = ((\widehat{V}^*,\widehat{A}^*),\widehat{\varphi}^*)\) 
obtained by contracting all arcs \((u,v)\)
where \(u\) has outdegree~1 (see Figure~\ref{fig:one:nested:x:networks}(c)).

\section{Rankings, time-stamp functions and equidistant $X$-cactuses}
\label{sub:sec:rankings}

\begin{figure}
\centering
\includegraphics[scale=0.9]{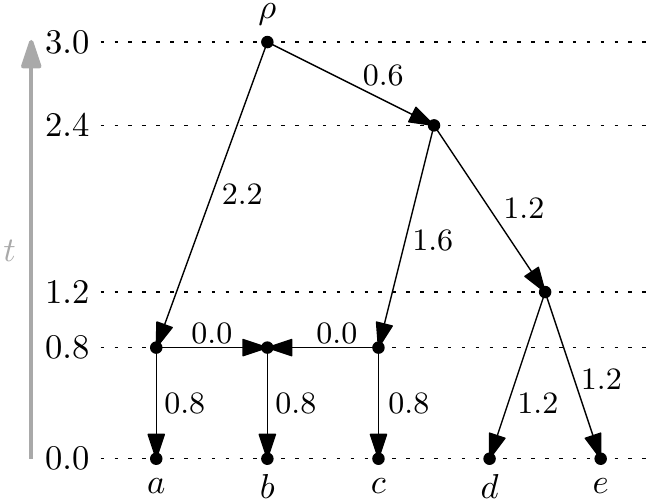}
\caption{A rooted \(X\)-cactus \(\mathcal{N}\)
on \(X=\{a,b,c,d,e\}\) with a time-stamp
function \(t\) on its vertices. For all vertices~\(v\) the value~\(t(v)\)
is given by the real number to the left of the horizontal line through~\(v\).
In addition, for each arc of \(\mathcal{N}\), the length of the arc induced by~\(t\) 
is given.}
\label{fig:equidistant:network}
\end{figure}

In this section, we consider rankings of the vertices of
rooted $X$-cactuses, which
are an important part of defining equidistant-cactus space.
It is convenient to start with the more general
concept of time-stamp functions, which also naturally
leads to the definition of equidistant $X$-cactuses.
A \emph{time-stamp function} on the vertices
in a rooted \(X\)-cactus \(\mathcal{N} = ((V,A),\varphi)\)
is a map \(t : V \rightarrow \mathbb{R}_{\geq 0}\) such that
\begin{itemize}
\setlength{\itemindent}{15pt}
\item[(TS1)]
\(t(v)=0\) for all \(v \in \varphi(X)\),
\item[(TS2)]
\(t(u) > t(v)\) for all arcs \((u,v)\) of \(\mathcal{N}\) with \(v\)
not a reticulation vertex, and
\item[(TS3)]
\(t(v) = t(p_1) = t(p_2)\) for all reticulation vertices \(v\)
of \(\mathcal{N}\) and its two parents \(p_1\) and \(p_2\).
\end{itemize}
An example of a time-stamp function
on the vertices of a rooted \(X\)-cactus
is given in Figure~\ref{fig:equidistant:network}.
Integer-valued time-stamp functions are also known as
\emph{temporal labelings} (see e.g.~\cite{BSS06}).
We call a rooted \(X\)-cactus~\(\mathcal{N}\) \emph{temporal}
if there exists a time-stamp function on the vertices of~\(\mathcal{N}\).
Note that not every rooted \(X\)-cactus
is temporal (for example, the rooted \(X\)-cactus
in Figure~\ref{fig:one:nested:x:networks}(a) is not temporal
because \(\varphi(X)\) contains an internal vertex that is
not a parent of a reticulation vertex).
The following lemma characterizes rooted \(X\)-cactuses that are temporal
(see also~\cite[Theorem~3]{BSS06} for a characterization
that applies to general rooted phylogenetic networks).

\begin{lemma}
A rooted \(X\)-cactus~\(\mathcal{N} = ((V,A),\varphi)\) is temporal
if and only if for all vertices \(u \in V\) the
following properties hold: 
\begin{itemize}
\item[(a)]
If \(u \in \varphi(X)\) then either \(u\) is a leaf or a parent of
a reticulation vertex that is a leaf.
\item[(b)]
If \(u\) has outdegree at least~2 then \(u\)
is not the parent of a reticulation vertex that is a leaf.
\item[(c)]
If \(u\) is the parent of a reticulation vertex \(v\) in a
reticulation cycle \(\{P,P'\}\) then neither of the directed paths
\(P\), \(P'\) consists of the single arc \((u,v)\).
\end{itemize}
\end{lemma}

\begin{proof}
First assume that \(\mathcal{N}\) is temporal. Consider
a time-stamp function \(t\) on the vertices of~\(\mathcal{N}\).
Assuming that~\(\mathcal{N}\)
contains a vertex~\(u\) that violates one of~(a)-(c) 
immediately yields a contradiction because then~\(t\) would
violate at least one of~(TS1)-(TS3).

Now assume that~(a)-(c) hold for all vertices of \(\mathcal{N}\).
We construct a time-stamp function \(t\) on the vertices of \(\mathcal{N}\)
by first putting \(t(v) = 0\) for all \(v \in \varphi(X)\).
In view of~(a) and~(b), this does not violate~(TS1)-(TS3).

Next, consider an internal vertex \(u\) that is not a reticulation
vertex and also not the parent of a reticulation vertex.
Assume that all children~\(w\) of~\(u\) have been assigned time-stamps~\(t(w)\).
Then we put \(t(u) = 1 + \max_{w \in ch(u)} t(w)\). Since
\(\mathcal{N}\) is acyclic this does not violate~(TS1)-(TS3).

Finally, consider an internal vertex \(u\) that is
a reticulation vertex. Let \(p_1\) and \(p_2\) denote the
two parents of \(u\) and assume that all vertices~\(w\) in
\[M = (ch(u) \cup ch(p_1) \cup ch(p_2)) \setminus \{u\}\]
have been assigned time-stamps~\(t(w)\). Then we put
\(t(u) = t(p_1) = t(p_2) = 1 + \max_{w \in M} t(w)\).
Since \(\mathcal{N}\) is acyclic and in view of~(c) this does not violate~(TS1)-(TS3).

Thus, our inductive construction yields a map \(t : V \rightarrow \mathbb{R}_{\geq 0}\)
for which~(TS1)-(TS3) hold.
\end{proof}

As indicated in Figure~\ref{fig:equidistant:network}, a time-stamp
function \(t\) on the vertices of a rooted \(X\)-cactus
\(\mathcal{N} = ((V,A),\varphi)\) induces non-negative lengths
on the arcs of~\(\mathcal{N}\)
by putting the length of arc \((u,v)\) to be \(t(u)-t(v)\). With these
arc lengths, all directed paths from a fixed vertex~\(u\) to 
a vertex \(w \in \varphi(X)\) have the same length, namely~\(t(u)\).
In view of this, we call an ordered pair \((\mathcal{N},t)\)
consisting of a rooted, temporal \(X\)-cactus~\(\mathcal{N}\)
and a time-stamp function~\(t\) on the vertices of~\(\mathcal{N}\)
an \emph{equidistant} \(X\)-cactus.
Thus, an equidistant \(X\)-cactus can be thought of as
a rooted, temporal \(X\)-cactus with specific arc lengths assigned,
whereas a rooted, temporal
\(X\)-cactus does not have any specific arc lengths assigned.

\begin{figure}
\centering
\includegraphics[scale=0.9]{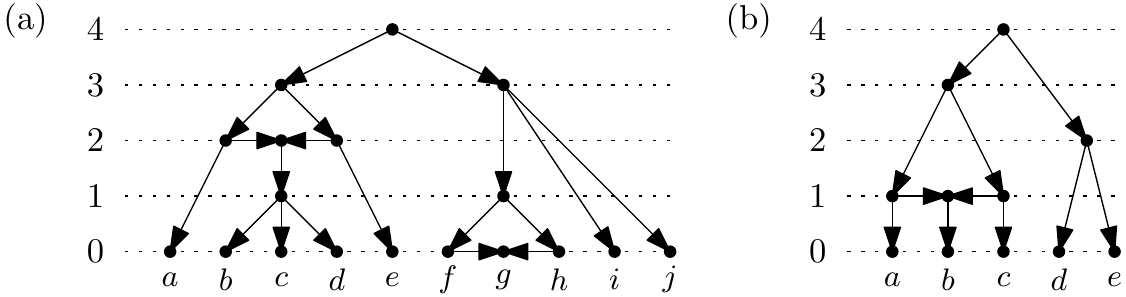}
\caption{(a) A ranking of size~4 of a rooted
\(X\)-cactus with \(X=\{a,b,c,\dots,j\}\).
Vertices of the same rank are drawn on the same
horizontal line.
(b) A ranking of a rooted, binary \(X\)-cactus
with \(X=\{a,b,c,d,e\}\). The ranking has size~4
which is the maximum size over all rooted, temporal
\(X\)-cactuses with $\lvert X \rvert = 5$.}
\label{fig:rankable:network}
\end{figure}

We conclude this section by shedding some more light on the
combinatorial structure of rooted, temporal \(X\)-cactuses.
The \emph{size}~\(\sigma(t)\) of a time-stamp function~\(t\)
on the vertices of a rooted, temporal \(X\)-cactus \(\mathcal{N} = ((V,A),\varphi)\)
is $\lvert t(V) \rvert - 1$. A \emph{ranking} of a rooted, temporal
\(X\)-cactus \(\mathcal{N} = ((V,A),\varphi)\)
is a time-stamp function \(r\) on the vertices of \(\mathcal{N}\) with
\(r(V) = \{0,1,2,\dots,\sigma(r)\}\).
See Figure~\ref{fig:rankable:network}(a) for an example.
Note that rankings as defined here are a particular 
type of temporal labeling and are more general than
the rankings considered in~\cite{BLS20a}.
The value \(r(v)\) assigned to vertex \(v\)
by the ranking~\(r\) will also be referred to as the
\emph{rank} of vertex~\(v\) if the ranking referred to is
clear from the context.  
A \emph{ranked} \(X\)-cactus \((\mathcal{N},r)\)
consists of a rooted, temporal \(X\)-cactus~\(\mathcal{N}\)
and a ranking~\(r\) of the vertices of~\(\mathcal{N}\).
The following lemma gives tight bounds on the size of rankings
of rooted, temporal \(X\)-cactuses (see Figure~\ref{fig:rankable:network}(b)
for an example). For its proof, we will use the fact that any 
rooted binary \(X\)-cactus can be transformed
into a rooted binary \(X\)-tree by deleting, for every
reticulation vertex~\(v\), one of the arcs \((p,v)\) from 
a parent~\(p\) of \(v\) to \(v\) and then suppressing the two
internal vertices \(v\) and \(p\).

\begin{lemma}
\label{lem:size:ranking}
Let \((\mathcal{N},r)\) be a ranked \(X\)-cactus. Then
we have \(0 \leq \sigma(r) \leq \lvert X \rvert - 1\). Moreover, 
\begin{itemize}
\item[(a)]
\(\sigma(r) = 0\) if and only if~\(\mathcal{N}\) consists of
a single vertex.
\item[(b)]
\(\sigma(r) = \lvert X \rvert -1\) if and only if~\(\mathcal{N}\) is a
rooted, binary \(X\)-cactus
and \(r(u) \neq r(v)\) for all distinct vertices \(u\) and \(v\)
unless \(u\) and \(v\) are both leaves of~\(\mathcal{N}\),
\(u\) is a parent of a reticulation vertex~\(v\), or
\(u\) and \(v\) are parents of the same reticulation vertex.
\end{itemize}
\end{lemma}

\begin{proof}
By definition, \(\sigma(r) \geq 0\). Moreover,
if the size of the ranking~\(r\) is precisely~0 then \(\mathcal{N}\)
must consist of a single leaf~\(v\) with \(r(v) = 0\) and all
elements of \(X\) are mapped by~\(\varphi\) to~\(v\).

To establish the upper bound, let \(i\) and \(k\) denote the
number of internal and reticulation vertices, respectively,
of the ranked \(X\)-cactus~\((\mathcal{N},r)\). 
By definition, \(\sigma(r) \leq (i - 2k)\).
Note that, for fixed~\(X\),
this expression can only be maximum if~\(\mathcal{N}\) is
a rooted, binary \(X\)-cactus,
because otherwise we can always increase~\(i\)
without increasing~\(k\). Hence, it suffices to show that for all
rooted, binary \(X\)-cactuses we have \(i - 2k = \lvert X \rvert - 1\).
Since, as described above, we can transform any such \(X\)-cactus
into a rooted binary \(X\)-tree, we immediately obtain this 
equation as a consequence of the well-known fact that
a rooted binary \(X\)-tree has $\lvert X \rvert - 1$ internal vertices
(see e.g. \cite[Sec.~2.1]{SS03a}).
\end{proof}

\section{Equidistant $X$-trees and $\tau$-space}
\label{sec:equidistant:x:trees}

In this section, we shall briefly recall the concept of an
orthant space (see e.g.~\cite[Sec.~6]{MOP15a}) and related concepts.
To illustrate the basic idea for constructing our orthant space
of equidistant-cactuses, we also consider the
simpler case of equidistant-trees (often called
ultrametric trees) and explain how the \(\tau\)-space of ultrametric trees
mentioned in the introduction arises as an orthant space.
This also yields an alternative
proof to the one presented in \cite{GD16} for the fact
that $\tau$-space is a \cat(0)-metric space.

\subsection{Orthant spaces}
\label{sub:sec:orthant:spaces}

An ordered pair \((M,\mathcal{F})\)
consisting of a family \(\mathcal{F}\) of non-empty subsets
of a finite non-empty set~\(M\)
is called an \emph{abstract simplicial complex} if \(A \in \mathcal{F}\) implies that
all non-empty subsets of~\(A\) are also contained in~\(\mathcal{F}\).
An abstract simplicial complex is a \emph{flag complex} if, for all
non-empty subsets \(A \subseteq M\) such that all 2-element subsets
of \(A\) are contained in \(\mathcal{F}\), we have \(A \in \mathcal{F}\).
For every map \(\omega : M \rightarrow \mathbb{R}_{\geq 0}\) we put
\(\supp(\omega) = \{x \in M : \omega(x) > 0\}\). 
The \emph{orthant space} associated with the abstract
simplicial complex \((M,\mathcal{F})\) is
\[\mathfrak{M}_{(M,\mathcal{F})} = \{\omega \in \mathbb{R}_{\geq 0}^M : 
\supp(\omega) \in \mathcal{F} \cup \{\emptyset\}\}.\]
A \emph{metric} \(D\) on a non-empty set~\(B\) is a map
\(D : B \times B \rightarrow \mathbb{R}_{\geq 0}\) such that 
\begin{itemize}
\item
\(D(x,y) = 0\) if and only if \(x=y\),
\item
\(D(x,y) = D(y,x)\), and
\item
\(D(x,z) \leq D(x,y) + D(y,z)\)
\end{itemize}
hold for all \(x,y,z \in B\). The ordered pair \((B,D)\) is 
called a \emph{metric space} and the elements of \(B\) are
called the \emph{points} of the metric space.
A metric \(D_{(M,\mathcal{F})}\) on the orthant space \(\mathfrak{M}_{(M,\mathcal{F})}\)
associated with the abstract simplicial complex \((M,\mathcal{F})\)
can be constructed as follows.
For every \(A \in \mathcal{F}\), the set
\[\mathfrak{O}(A) = \{\omega \in \mathfrak{M}_{(M,\mathcal{F})} : \supp(\omega) \subseteq A\}\]
is called an \emph{orthant} of~\(\mathfrak{M}_{(M,\mathcal{F})}\).
For all \(\omega,\omega' \in \mathfrak{M}_{(M,\mathcal{F})}\) such that there exists
an orthant \(\mathfrak{O}\) of~\(\mathfrak{M}_{(M,\mathcal{F})}\) with
\(\{\omega,\omega'\} \subseteq \mathfrak{O}\) we put 
\[D_{(M,\mathcal{F})}(\omega,\omega') = \sqrt{\sum_{x \in M} 
(\omega(x) - \omega'(x))^2}.\]
Then, for all \(\omega,\omega' \in \mathfrak{M}_{(M,\mathcal{F})}\)
such that there is no orthant \(\mathfrak{O}\) of~\(\mathfrak{M}_{(M,\mathcal{F})}\)
that contains both \(\omega\) and \(\omega'\) we consider
finite \emph{segmented paths} from \(\omega\) to \(\omega'\).
These are sequences \(\omega_0,\omega_1,\omega_2,\dots,\omega_k\) of
elements in \(\mathfrak{M}_{(M,\mathcal{F})}\) such that 
\(\omega=\omega_0\), \(\omega'=\omega_k\) and,
for all \(i \in \{1,2,\dots,k\}\),
there exists some orthant \(\mathfrak{O}_i\) of \(\mathfrak{M}_{(M,\mathcal{F})}\)
that contains both \(\omega_{i-1}\) and \(\omega_i\). 
The \emph{length} of such a segmented path is
\(\sum_{i=1}^k D_{(M,\mathcal{F})}(\omega_{i-1},\omega_{i})\).
Note that at least one such segmented path
always exists in view of the fact that
all orthants of~\(\mathfrak{M}_{(M,\mathcal{F})}\) contain the point
\(\omega\) with \(\supp(\omega) = \emptyset\),
called the \emph{origin} of~\(\mathfrak{M}_{(M,\mathcal{F})}\).
We define \(D_{(M,\mathcal{F})}(\omega,\omega')\) to be the 
infimum of the length of all segmented paths from~\(\omega\) to~\(\omega'\).
It is known (see \cite[Sec.~6]{MOP15a}) that this construction
yields a metric space~\((\mathfrak{M}_{(M,\mathcal{F})},D_{(M,\mathcal{F})})\).

Next, we describe a useful property that
the metric space~\((\mathfrak{M}_{(M,\mathcal{F})},D_{(M,\mathcal{F})})\) may have.
A \emph{geodesic path} between the points \(p\) and \(q\) in
a metric space \((B,D)\) is a map
\(\gamma : [0,\ell] \rightarrow B\), for some \(\ell \geq 0\),
with \(\gamma(0)=p\), \(\gamma(\ell) = q\) and
\(D(\gamma(t_1),\gamma(t_2)) = \vert t_1 - t_2 \rvert\) for all \(t_1,t_2 \in [0,\ell]\).
A metric space \((B,D)\) is \emph{geodesic} if there exists a geodesic path
between~\(p\) and~\(q\) for all \(p,q \in B\). A geodesic metric space \((B,D)\) is
a \(\cat(0)\)-metric space if and only if (see e.g.~\cite[p.~163]{BH99a})
\[(D(p,q))^2 + (D(p,r))^2 \geq 2(D(m,p))^2 + (D(q,r))^2/2\]
holds for all \(p,q,r \in B\)
and all \(m \in B\) with \(D(q,m) = D(r,m) = D(q,r)/2\).
\(\cat(0)\)-metric spaces arise in many applications (see e.g.~\cite{AM20a}).
They have the important property that geodesic paths are
unique~\cite[Proposition~1.4, p.~160]{BH99a}. 
It follows from a result in \cite{gromov1987hyperbolic}
that the orthant space \((\mathfrak{M}_{(M,\mathcal{F})},D_{(M,\mathcal{F})})\) is a
\(\cat(0)\)-metric space if and only if~\(\mathcal{F}\) is a
flag complex (see also \cite[Proposition~6.14]{MOP15a}).
Furthermore, geodesic paths can be computed in
polynomial time in \(\cat(0)\)-orthant spaces \cite[Corollary~6.19]{MOP15a}.

\subsection{$\tau$-space revisited}
\label{sub:sec:tree:space}

To describe how the \(\tau\)-space of ultrametric trees arises as an orthant space,
we start with a suitably defined abstract simplicial complex. 
A \emph{partition} of~\(X\) is a set~\(\mathcal{P}\) of
non-empty and pairwise disjoint subsets of~\(X\) with
\(X = \bigcup_{A \in \mathcal{P}} A\).
We denote the set of all partitions of~\(X\) by~\(\mathfrak{B}(X)\) and
define a binary relation~\(\sqsubseteq\) on~\(\mathfrak{B}(X)\) by
putting \(\mathcal{P}_1 \sqsubseteq \mathcal{P}_2\) if for all
\(A_1 \in \mathcal{P}_1 \) there exists some \(A_2 \in \mathcal{P}_2\) with
\(A_1 \subseteq A_2\). Intuitively, this means that the
partition~\(\mathcal{P}_1\) refines the partition~\(\mathcal{P}_2\).
It is well-known that~\(\sqsubseteq\) is a partial ordering.
Note that the partial ordering~\(\sqsubseteq\) is induced by
the partial ordering~\(\subseteq\) on the subsets of~\(X\). 

\begin{figure}
\centering
\includegraphics[scale=1.0]{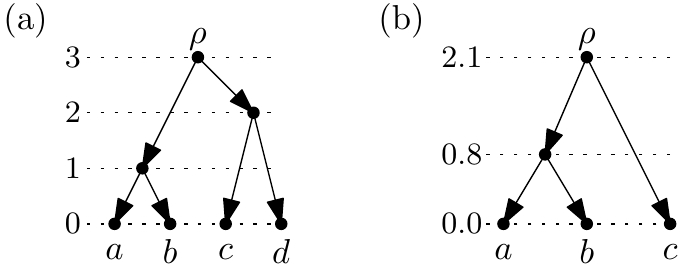}
\caption{(a) A ranked \(X\)-tree with \(X=\{a,b,c,d\}\). Any cut 
along one of the dotted horizontal lines yields a partition of \(X\)
(for example, the dotted line labeled with~1 yields the 
partition $\{\{a,b\},\{c\},\{d\}\}$).
(b) An equidistant \(X\)-tree with \(X=\{a,b,c\}\).}
\label{fig:examples:trees}
\end{figure}

Every ranked \(X\)-tree with a ranking of size~\(\sigma\) gives
rise to a sequence
\[\mathcal{P}_0 \sqsubseteq \mathcal{P}_1 \sqsubseteq \dots
\sqsubseteq  \mathcal{P}_{\sigma} = \{X\}\]
of partitions of~\(X\). In Figure~\ref{fig:examples:trees}(a)
we depict a rooted \(X\)-tree with a ranking of size \(\sigma = 3\)
that gives rise to the sequence
\[\{\{a\},\{b\},\{c\},\{d\}\} \sqsubseteq
\{\{a,b\},\{c\},\{d\}\}  \sqsubseteq
\{\{a,b\},\{c,d\}\}  \sqsubseteq
\{\{a,b,c,d\}\}\]
(see also Section~\ref{sec:set:pair:systems} where we
formally define how the partitions arise more generally
for ranked \(X\)-cactuses).
The crucial fact is that this sequence encodes
the ranked $X$-tree. More formally,
as we shall prove as a consequence of our results for
general ranked \(X\)-cactuses
in Corollary~\ref{cor:chains:rooted:x:trees}, we have:

\begin{theorem}
\label{prop:pairwise:trees}
There is a one-to-one correspondence between 
(isomorphism classes of) ranked \(X\)-trees and
subsets of \(\mathfrak{B}(X)\) that contain~\(\{X\}\)
and that consist of partitions of~\(X\) which are pairwise
comparable with respect to the partial ordering~\(\sqsubseteq\).
\end{theorem}

To obtain \(\tau\)-space as an orthant space, we
consider the abstract simplicial complex
\((\mathfrak{B}^{\circ}(X),\mathcal{F}(\sqsubseteq))\) with
\(\mathfrak{B}^{\circ}(X) = \mathfrak{B}(X) - \{X\}\) and
\(\mathcal{F}(\sqsubseteq)\) containing all non-empty subsets
of \(\mathfrak{B}^{\circ}(X)\) whose elements are pairwise
comparable with respect to~\(\sqsubseteq\). 
It follows immediately that
\((\mathfrak{B}^{\circ}(X),\mathcal{F}(\sqsubseteq))\) is 
a flag complex. Note that, more generally, we can associate
an abstract simplicial complex that is a flag complex
to any partial ordering in an analogous way;
for this reason such a complex is known as an \emph{order complex}
(see e.g.~\cite[p.~248]{WW92a}).

\begin{figure}
\centering
\includegraphics[scale=0.9]{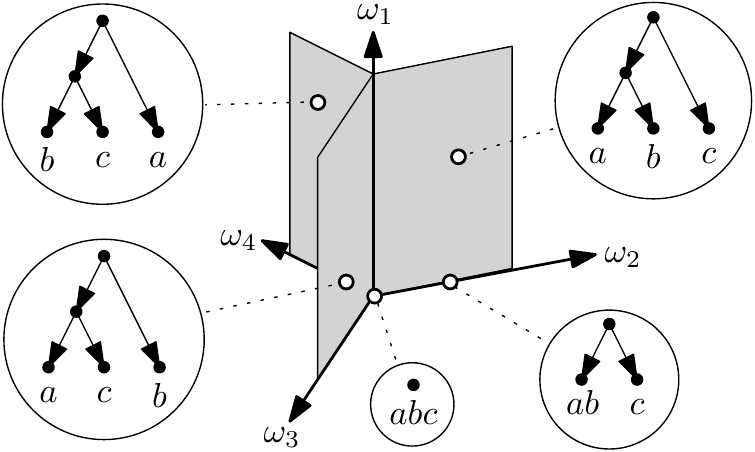}
\caption{The orthant space
\(\mathfrak{M}_{(\mathfrak{B}^{\circ}(X),\mathcal{F}(\sqsubseteq))}\) for \(X=\{a,b,c\}\).
By construction,
each axis represents a partition of \(X\) distinct from~\(\{X\}\).
The axes labeled~\(\omega_1\) and~\(\omega_2\), for example, represent
the partitions \(\{\{a\},\{b\},\{c\}\}\) and \(\{\{a,b\},\{c\}\}\), respectively.
The three 2-dimensional orthants are drawn shaded.
All points in the interior of these 2-dimensional orthants correspond to the same 
isomorphism class of binary ranked \(X\)-trees. The
rankings for them are not shown because they are unique.
Points on the axes correspond to non-binary ranked \(X\)-trees.
The origin corresponds to the ranked \(X\)-tree
that consists of a single vertex.}
\label{fig:orthants:trees:3:elements}
\end{figure}

In Figure~\ref{fig:orthants:trees:3:elements}, we illustrate
the orthant space
\(\mathfrak{M}_{(\mathfrak{B}^{\circ}(X),\mathcal{F}(\sqsubseteq))}\)
of equidistant $X$-trees for $X=\{a,b,c\}$
(see Figure~\ref{fig:orthants:networks:3:elements}
for an analogous drawing of the resulting orthant space
of equidistant \(X\)-cactuses).
Note that, by construction, the 
coordinates of a point in any orthant are obtained as
differences between consecutive time stamps in the
equidistant $X$-tree that corresponds to the point.
The equidistant \(X\)-tree in Figure~\ref{fig:examples:trees}(b),
for example, corresponds to the point
\((\omega_1,\omega_2,\omega_3,\omega_4) = (0.8,1.3,0,0)\).
More generally, it follows by Theorem~\ref{prop:pairwise:trees}
that the elements in \(\mathfrak{M}_{(\mathfrak{B}^{\circ}(X),\mathcal{F}(\sqsubseteq))}\)
are in one-to-one correspondence with equidistant $X$-trees.
Moreover, since \((\mathfrak{B}^{\circ}(X),\mathcal{F}(\sqsubseteq))\) is 
a flag complex, it follows, as mentioned in Section~\ref{sub:sec:tree:space},
that the resulting metric space 
\((\mathfrak{M}_{(\mathfrak{B}^{\circ}(X),\mathcal{F}(\sqsubseteq))},D_{(\mathfrak{B}^{\circ}(X),\mathcal{F}(\sqsubseteq))})\) is~\cat(0).
We remark that, by construction, \(\mathfrak{M}_{(\mathfrak{B}^{\circ}(X),\mathcal{F}(\sqsubseteq))}\) is precisely $\tau$-space, and so
we obtain an alternative 
proof to the one presented in \cite{GD16} that $\tau$-space
is a \cat(0)-metric space.

Before proceeding,
we note that in \cite{HSS21a} 
the problem of when a partition of~\(X\) is compatible with
a rooted phylogenetic \(X\)-tree is studied. This includes, as a special case,
the situation where the vertices of the tree can be ranked in such a way that
the partition is among those associated with the resulting ranked \(X\)-tree.
In addition, in \cite{AK06a} a space, called the \emph{Bergman fan}
of the matroid of the complete graph with vertex set~\(X\) is studied.
This space is a polyhedral fan ans its
points are also in one-to-one correspondence with
equidistant \(X\)-trees. Although not an orthant space,
its cones are in one-to-one correspondence with the orthants of
\(\mathfrak{M}_{(\mathfrak{B}^{\circ}(X),\mathcal{F}(\sqsubseteq))}\).

\section{An encoding for ranked $X$-cactuses}\label{sec:encoding}

To help the reader navigate the remaining sections of this paper, we 
now briefly summarize how we  shall construct the
equidistant-cactus space $\mathfrak{N}(X)$
by applying an analogue of the process described in
Section~\ref{sub:sec:tree:space}.

We shall begin by introducing the
concept of a \emph{polestar system}
on the set~$X$, which is a collection
of ordered pairs of subsets of~\(X\), or
set pair system for short, with certain properties.
As we shall see in Section~\ref{sec:partition:like:set:pair:systems}, polestar systems can be 
associated to ranked $X$-cactuses in a similar way
how partitions can be associated to ranked $X$-trees. 
We shall also define a binary relation
\(\preceq\) on general set pair systems, and, 
in Section~\ref{sec:poset:p:x}, we will show that \(\preceq\) 
yields a partial ordering on the set $\mathfrak{P}(X)$ of 
polestar systems on $X$. 
In Section~\ref{sec:encoding:ranked:x:network}, we then prove an analogue of
Theorem~\ref{prop:pairwise:trees}, namely, we show
that ranked $X$-cactuses are in one-to-one 
correspondence with subsets of $\mathfrak{P}(X)$ that contain the maximum
element relative to the ordering $\preceq$ and
that are pairwise comparable with respect to~\(\preceq\).
In other words, we obtain an
encoding of ranked \(X\)-cactuses in terms
of certain collections of polestar systems. 
In Section~\ref{sec:conclusion}, we conclude by constructing
the network space $\mathfrak{N}(X)$ as the orthant space associated to
the order complex of the poset $(\mathfrak{P}(X),\preceq)$.

\subsection{Set pair systems}
\label{sec:set:pair:systems}

Before introducing polestar systems, we recall
the concept of a set pair system introduced in~\cite{HMS21a}.
To this end, we say that a vertex $u$ in a rooted DAG~\(N\)
is a \emph{descendant} of a vertex $v$
if there exists a directed path from the root of \(N\) to \(u\) that
contains \(v\). 
A descendant \(u\) of \(v\) is a \emph{strict descendant} if
every directed path from the root to $u$ contains $v$. Otherwise \(u\) is called
a \emph{non-strict descendant} of \(v\).
Now, given a rooted \(X\)-cactus $\mathcal{N} = ((V,A),\varphi)$
and a vertex $u \in V$,
let \(C(u)\) be the set of those \(x \in X\) with \(\varphi(x)\) a
descendant of $u$,
$S(u)$ the set of those \(x \in X\) with \(\varphi(x)\) a strict
descendant of $u$ and $H(u)$
the set of those \(x \in X\) with \(\varphi(x)\) a non-strict
descendant of $u$ in $X$.
For every vertex \(u\) of \(\mathcal{N}\) we call \((S(u),H(u))\)
the \emph{set pair} associated to~\(u\) and put
\[\mathcal{S}(\mathcal{N}) = \{(S(u),H(u)) : u \in V\}.\]
For later reference, we state some immediate consequences of
the definition of the set pairs in \(\mathcal{S}(\mathcal{N})\)
for a rooted \(X\)-cactus~\(\mathcal{N}\) (see also~\cite{HMS21a}
where these properties have been considered in the context of
the slightly more restrictive 1-nested phylogenetic networks):
\begin{itemize}
\setlength{\itemindent}{15pt}
\item[(SH1)]
For all vertices \(u\) of \(\mathcal{N}\), we have
\(S(u) \cap H(u) = \emptyset\),
\(S(u) \cup H(u) = C(u)\) and
$S(u)$ is always non-empty while $H(u)$ may be empty.
\item[(SH2)]
If \((S(u),H(u)) = (S(v),H(v))\) for two distinct vertices
\(u\) and \(v\) of \(\mathcal{N}\) then one of these vertices,
say~\(u\), is a reticulation vertex 
with outdegree~1 and \(v\) is the single child of~\(u\).
Note that this situation cannot occur if~\(\mathcal{N}\) is compressed.
\item[(SH3)]
Let \(C\) be the set of vertices in a reticulation cycle of \(\mathcal{N}\)
where \(u\) and \(v\) are the common start and end vertex, respectively,
of the two directed paths that form the reticulation cycle.
Then we have \(H(w) = S(v)\)
if \(w \in C - \{u,v\}\) and, for all other vertices \(w'\) of \(\mathcal{N}\),
we have \(H(w') \neq S(v)\).
\end{itemize}

Now, given a ranked \(X\)-cactus \((\mathcal{N}=((V,A),\varphi),r)\)
we collect, for every \(i \in \{0,1,2,\dots,\sigma(r)\}\),
in \(\mathcal{S}_i(\mathcal{N})\) first those set
pairs from~\(\mathcal{S}(\mathcal{N})\) that correspond to vertices
of rank at most~\(i\) and whose parents (if any) have rank strictly larger
than~\(i\). We then add some further set pairs that essentially help to keep
track of the fact that some of the vertices involved are in a reticulation cycle.
More formally, we define \(V_i\) to be the set that consists of 
all vertices \(u \in V\) with
\(r(u) \leq i\) and
\(r(p) > i\) for all parents~\(p\) of~\(u\).
Note that, in view of~(TS3), \(V_i\)~does not contain any reticulation vertices.
Thus, all \(u \in V_i\) have at most one parent. Then we put
\[\mathcal{S}_i(\mathcal{N}) = \{(S(u),H(u)) : u \in V_i\} \cup
\{(H(u),\emptyset) : u \in V_i, \ H(u) \neq \emptyset\}.\]
Note that we always have \(\mathcal{S}_{\sigma(r)} = \{(X,\emptyset)\}\).
For the rooted \(X\)-cactus \(\mathcal{N}\) in Figure~\ref{fig:rankable:network}(a),
for example, we obtain:
{\small
\begin{align*}
\mathcal{S}_4(\mathcal{N}) &= \{(\{a,b,c,d,e,f,g,h,i,j\},\emptyset)\}\\
\mathcal{S}_3(\mathcal{N}) &= \{(\{a,b,c,d,e\},\emptyset),(\{f,g,h,i,j\},\emptyset)\}\\
\mathcal{S}_2(\mathcal{N}) &= \{(\{a\},\{b,c,d\}),(\{b,c,d\},\emptyset),(\{e\},\{b,c,d\}),
(\{f,g,h\},\emptyset),(\{i\},\emptyset),(\{j\},\emptyset)\}\\
\mathcal{S}_1(\mathcal{N}) &= \{(\{a\},\emptyset),(\{b,c,d\},\emptyset),(\{e\},\emptyset),
(\{f,g,h\},\emptyset),(\{i\},\emptyset),(\{j\},\emptyset)\}\\
\mathcal{S}_0(\mathcal{N}) &= \{(\{x\},\emptyset): x \in \{a,b,c,d,e,g,i,j\}\} \cup
\{(\{f\},\{g\}),(\{h\},\{g\})\}
\end{align*}}

A collection of ordered pairs $(S,H)$ of subsets of \(X\) 
such that  \(S \neq \emptyset\)
and \(S \cap H = \emptyset\) is called a  \emph{set pair system} on~\(X\).
Note that, by construction,
the sets \(\mathcal{S}(\mathcal{N})\) and \(\mathcal{S}_i(\mathcal{N})\),
\(0 \leq i \leq \sigma(r)\), associated
with a ranked \(X\)-cactus \((\mathcal{N},r)\) are non-empty 
set-pair systems.

It is shown in \cite{HMS21a}
that, for any set pair system \(\mathcal{S}\) on \(X\), 
we obtain a partial ordering \(\leq\) on the set pairs in \(\mathcal{S}\)
by putting \((S_1,H_1) \leq (S_2,H_2)\)
if either \((S_1,H_1) = (S_2,H_2)\) or \((S_1,H_1) \neq (S_2,H_2)\) and one the
following holds:
\begin{itemize}
\item
\(S_1 \cup H_1 \subseteq S_2\)
\item
\(S_1 \cup H_1 \subseteq H_2\)
\item
\(S_1 \subsetneq S_2\) and \(H_1 = H_2 \neq \emptyset\)
\end{itemize}
We write \((S_1,H_1) < (S_2,H_2)\) if \((S_1,H_1) \leq (S_2,H_2)\)
and the set pairs \((S_1,H_1)\) and \((S_2,H_2)\) are distinct.
The partial ordering \(\leq\) on set pairs
was defined in such a way that
we have \((S(u),H(u)) \leq (S(v),H(v))\) for two vertices~\(u\)
and~\(v\) in a rooted \(X\)-cactus if and only if
\(u\) is a descendant of \(v\)
(see the proof Theorem~5 in \cite{HMS21a}).

We use the partial ordering \(\leq\) on set pairs to define a
binary relation~\(\preceq\) on set pair systems. More precisely,
for set pair systems~\(\mathcal{S}_1\) and~\(\mathcal{S}_2\) on~\(X\)
we put \(\mathcal{S}_1 \preceq \mathcal{S}_2\) if 
\begin{itemize}
\setlength{\itemindent}{15pt}
\item[(SP1)]
for all \((S_1,H_1) \in \mathcal{S}_1\) there exists some
\((S_2,H_2) \in \mathcal{S}_2\) with \((S_1,H_1) \leq (S_2,H_2)\), and
\item[(SP2)]
for all \((S_2,H_2) \in \mathcal{S}_2\) with \(H_2 \neq \emptyset\),
if there exists some \((S_1,H_1) \in \mathcal{S}_1\) with
\(H_1 = H_2\), then there exists such a \((S_1,H_1)\) with
\((S_1,H_1) \leq (S_2,H_2)\).
\end{itemize}
Again, we write \(\mathcal{S}_1 \prec \mathcal{S}_2\) if
\(\mathcal{S}_1 \preceq \mathcal{S}_2\) and \(\mathcal{S}_1 \neq \mathcal{S}_2\).
We remark that~(SP1) captures the basic idea from Section~\ref{sub:sec:tree:space}
that the partial ordering
\(\leq\) on set pairs induces a suitable binary relation on set pair
systems (in analogy to how the partial ordering \(\subseteq\) induced the
binary relation~\(\sqsubseteq\)).
(SP2) is an additional technical requirement that will be
crucial in our encoding of ranked \(X\)-cactuses.

The relation \(\preceq\) is, in general, not a partial ordering
on the set pair systems on a fixed set \(X\) because it might neither
be antisymmetric nor transitive. For the set pair systems
associated with a ranked \(X\)-cactus, however, the following holds.

\begin{lemma}
\label{lem:ranking:equals:partial:ordering}
Let \((\mathcal{N},r)\) be a ranked \(X\)-cactus.
Then we have \(\mathcal{S}_i(\mathcal{N}) \prec \mathcal{S}_j(\mathcal{N})\)
for all \(0 \leq i < j \leq \sigma(r)\) and
\(\mathcal{S}_{\sigma(r)}(\mathcal{N}) = \{(\{X\},\emptyset)\}\).
\end{lemma}

\begin{proof}
As noted earlier in this section,
\(\mathcal{S}_{\sigma(r)}(\mathcal{N}) = \{(\{X\},\emptyset)\}\)
follows immediately from the definition of the set pair system
\(\mathcal{S}_{\sigma(r)}(\mathcal{N})\).
Consider \(0 \leq i < j \leq \sigma(r)\).
We first show that \(\mathcal{S}_i(\mathcal{N}) \preceq \mathcal{S}_j(\mathcal{N})\).
So, consider \((S,H) \in \mathcal{S}_i(\mathcal{N})\).
By definition of \(\mathcal{S}_i(\mathcal{N})\),
there must exist a vertex \(v\) in \(\mathcal{N}\) with \(r(v) \leq i\),
\(r(p) > i\) for all parents~\(p\) of~\(v\), and
either \((S,H) = (S(v),H(v))\) or \((S,H) = (H(v),\emptyset)\).
Consider a
directed path from the root of \(\mathcal{N}\) to~\(v\).
On this path there must exist a vertex \(u\) with
\(r(u) \leq j\) and \(r(p) > j\) for all parents~\(p\) of \(u\).
This implies that \((S(u),H(u)) \in \mathcal{S}_j(\mathcal{N})\).
Moreover, in view of the fact that \(u\) lies on a directed
path from the root of \(\mathcal{N}\) to~\(v\),
we must have \((S,H) \leq (S(v),H(v)) \leq (S(u),H(u))\),
as required by~(SP1).

To establish that also~(SP2) is satisfied
for \(\mathcal{S}_i(\mathcal{N})\) and \(\mathcal{S}_j(\mathcal{N})\), consider
\((S,H) \in \mathcal{S}_j(\mathcal{N})\) with \(H \neq \emptyset\).
By definition of \(\mathcal{S}_j(\mathcal{N})\),
there must exist a vertex \(u\) in \(\mathcal{N}\) with 
\((S,H) = (S(u),H(u))\), \(r(u) \leq j\) and
\(r(p) > j\) for all parents~\(p\) of~\(u\). Now, if there
exists some \((S',H') \in \mathcal{S}_i(\mathcal{N})\) with \(H'=H\)
then there exists some vertex \(v\) in \(\mathcal{N}\) with
\((S',H') = (S',H) = (S(v),H(v))\),
\(r(v) \leq i\)
and \(r(p) > i\) for all parents~\(p\) of~\(v\).
This implies that \(u\) and \(v\) must be vertices in the
same reticulation cycle of \(\mathcal{N}\). Moreover, we
can choose \(v\) such that \(v\)
is a descendant of \(u\), implying that
\((S',H') = (S(v),H(v)) \leq (S(u),H(u)) = (S,H)\),
as required.

It remains to show that \(\mathcal{S}_i(\mathcal{N}) \neq \mathcal{S}_j(\mathcal{N})\).
By the definition of a ranked \(X\)-cactus, there must exist a vertex \(u \in V\)
with \(r(u) = j\). Without loss of generality we may assume that \(u\) is
not a reticulation vertex.  
If \((S(u),H(u)) \not \in \mathcal{S}_i(\mathcal{N})\) we are
done. So, assume for a contradiction
that \((S(u),H(u)) \in \mathcal{S}_i(\mathcal{N})\). In view
of \(i < j\) we have \(u \not \in V_i\).
Thus, there exists some \(v \neq u\)
in \(V_i\) such that either (i) \(H(v) \neq \emptyset\) and \((S(u),H(u)) = (H(v),\emptyset)\)
or (ii) \((S(u),H(u)) = (S(v),H(v))\).
If Case~(i) holds then, in view of (SH3), \(v\) must be a vertex in a reticulation cycle
with end vertex~\(u'\) and \((S(u'),H(u')) = (H(v),\emptyset) = (S(u),H(u))\).
Since \(u\) is not a reticulation vertex, it follows, by (SH2),
that \(u\) is the single child of \(u'\).
Consequently, \(i=r(v) > r(u) = j\), a contradiction. Similarly, 
if Case~(ii) holds then, again by (SH2), it follows that \(u\) is a reticulation
vertex and~\(v\) is the single child of~\(u\), a contradiction.
\end{proof}

\subsection{Polestar systems}
\label{sec:partition:like:set:pair:systems}

A set pair system \(\mathcal{S}\) on \(X\) is \emph{partition-like}~if
\begin{itemize}
\setlength{\itemindent}{15pt}
\item[(PL1)]
\(\mathcal{P}(\mathcal{S}) = \{S : (S,H) \in \mathcal{S}\}\)
is a partition of~\(X\),
\item[(PL2)]
for all \((S,H),(S',H') \in \mathcal{S}\) with \((S,H) \neq (S',H')\)
we have \(S \neq S'\), and
\item[(PL3)]
for all \((S,H) \in \mathcal{S}\) with \(H \neq \emptyset\) we have
\((H,\emptyset) \in \mathcal{S}\) and
there exists precisely one \((S',H') \in \mathcal{S}\)
with \((S',H') \neq (S,H)\) and \(H=H'\).
\end{itemize}
A \underline{p}artiti\underline{o}n-\underline{l}ik\underline{e}
\underline{s}e\underline{t} p\underline{a}i\underline{r} system
is called a \emph{polestar system},
for short.
In addition, we define
\(\mathcal{H}(\mathcal{S}) = \{H : (S,H) \in \mathcal{S}, H \neq \emptyset\}\).
Note that~(PL2) implies that $\lvert \mathcal{S} \rvert = \lvert \mathcal{P}(\mathcal{S}) \rvert$.

\begin{lemma}
\label{lem:ranked:networks:yield:partition:like}
Let \((\mathcal{N},r)\) be a ranked \(X\)-cactus. 
Then \(\mathcal{S}_i(\mathcal{N})\) is a polestar system
for all \(0 \leq i \leq \sigma(r)\).
\end{lemma}

\begin{proof}
Fix some \(i \in \{0,1,\dots,\sigma(r)\}\) and consider
two distinct vertices \(u_1, u_2 \in V_i\).
Put \((S_k,H_k) = (S(u_k),H(u_k))\), \(k \in \{1,2\}\).
Recall from the definition of the set~\(V_i\) that
both \(u_1\) and \(u_2\) have rank at most~\(i\) while
the ranks of their parents are strictly larger than~\(i\).
Thus, up to switching the roles of \(u_1\) and \(u_2\),
one of the following must hold:
\begin{itemize}
\item
Neither of \(u_1\) and \(u_2\) is a descendant of the other
and there is no reticulation cycle in \(\mathcal{N}\) that
contains both \(u_1\) and \(u_2\). Consequently, 
\((S(u_1) \cup H(u_1)) \cap (S(u_2) \cup H(u_2)) = \emptyset\).
Thus, the sets \(S(u_1)\), \(H(u_1)\), \(S(u_2)\) and
\(H(u_2)\) are pairwise disjoint.
\item
Both \(u_1\) and \(u_2\) are contained in the
same reticulation cycle in \(\mathcal{N}\) but neither
is a descendant of the other.
Consequently, \(H(u_1) = H(u_2) = H \neq \emptyset\)
and the sets \(S(u_1)\), \(S(u_2)\) and \(H\) are pairwise
disjoint.
\end{itemize}
It follows from this case analysis that (PL1) and (PL2)
hold for \(\mathcal{S}_i(\mathcal{N})\).

To see that also (PL3)
holds, consider a set pair \((S,H) \in \mathcal{S}_i(\mathcal{N})\)
with \(H \neq \emptyset\). By the definition of
\(\mathcal{S}_i(\mathcal{N})\) there must exist
a vertex \(u\) in \(\mathcal{N}\) with
\((S(u),H(u)) = (S,H)\) such that
\(r(u) \leq i\) and \(r(p) > i\) for all parents \(p\)
of~\(u\). In view of \(H(u) = H \neq \emptyset\),
vertex \(u\) must be contained in a reticulation cycle~\(\mathcal{C}\)
but cannot be the common start or the common end vertex
of the two directed paths that form~\(\mathcal{C}\).
Note that~\(\mathcal{C}\) contains a unique
vertex \(v \neq u\) with \(r(v) \leq i\) and
\(r(p) > i\) for all parents~\(p\) of \(v\).
Moreover, \(v\) cannot be
the common start or the common end vertex
of the two directed paths that form~\(\mathcal{C}\).
Since \(u\) and \(v\) are both contained in~\(\mathcal{C}\),
we have \(H(u) = H(v) = H\). Moreover, by (SH3),
there are no other vertices
\(w\) in \(\mathcal{N}\) with \(H(w) = H\), \(r(w) \leq i\) and
\(r(p) > i\) for all parents \(p\) of~\(w\). Finally, by construction,
we also have \((H,\emptyset) = (H(u),\emptyset) \in \mathcal{S}_i(\mathcal{N})\).
\end{proof}

We denote by~\(\mathfrak{P}(X)\) the set of
polestar systems on the set~\(X\).
Note that, even for the set pair systems in~\(\mathfrak{P}(X)\),
(SP1)~in the definition of the binary relation~\(\preceq\)
does not imply~(SP2), as can be seen from the set pair systems
\begin{align*}
\mathcal{S}_1 &= \{(\{a\},\{b\}),(\{b\},\emptyset),(\{c\},\{b\}),(\{d\},\emptyset)\}
\ \text{and}\\
\mathcal{S}_2 &= \{(\{a,c\},\{b\}),(\{b\},\emptyset),(\{d\},\{b\})\}
\end{align*}
on \(X=\{a,b,c,d\}\) which satisfy (PL1)-(PL3) and (SP1) but not~(SP2).

We conclude this section with two technical lemmas stating
some properties of
the relations \(\leq\) and \(\preceq\) that will
be used in Sections~\ref{sec:poset:p:x} and~\ref{sec:encoding:ranked:x:network}.
In particular, Lemma~\ref{lem:technical:ordering:consistent}
establishes that, up to a specific exception,
distinct set pairs within a single polestar system are incomparable
with respect to the partial ordering \(\leq\) and
the binary relations \(\leq\) and \(\preceq\) are consistent.
In our encoding of ranked \(X\)-cactuses this exception
corresponds to the set pairs associated with reticulation
vertices.

\begin{lemma}
\label{lem:technical:ordering:consistent}
Let \(\mathcal{S}_1,\mathcal{S}_2 \in \mathfrak{P}(X)\) with
\(\mathcal{S}_1 \preceq \mathcal{S}_2\). Then,
for all \((S_1,H_1) \in \mathcal{S}_1\) and \((S_2,H_2) \in \mathcal{S}_2\),
\((S_2,H_2) < (S_1,H_1)\) implies
\((S_2,H_2) \in \mathcal{S}_1\), \(H_2 = \emptyset\) and \(H_1 = S_2\). 
\end{lemma}

\begin{proof}
First, consider the case \(\mathcal{S}_1 = \mathcal{S}_2 = \mathcal{S}\).
Let \((S_1,H_1), (S_2,H_2) \in \mathcal{S}\) with \((S_2,H_2) < (S_1,H_1)\).
Assume for a contradiction that \(H_2 \neq \emptyset\). Then,
in view of (PL1)-(PL3), none of
\(S_2 \cup H_2 \subseteq S_1\), \(S_2 \cup H_2 \subseteq H_1\) and
\(S_2 \subsetneq S_1\) can hold, in contradiction to \((S_2,H_2) < (S_1,H_1)\).
Thus, we must have \(H_2 = \emptyset\).
Consequently, \(S_2 \subseteq H_1\),
and, therefore, \(S_2=H_1\), as required.

Next consider the case \(\mathcal{S}_1 \prec \mathcal{S}_2\).
Let \((S_1,H_1) \in \mathcal{S}_1\) and \((S_2,H_2) \in \mathcal{S}_2\)
with \((S_2,H_2) < (S_1,H_1)\). In view of \(\mathcal{S}_1 \prec \mathcal{S}_2\),
there must exist some \((S_2',H_2') \in \mathcal{S}_2\) with
\((S_1,H_1) \leq (S_2',H_2')\). By the transitivity of \(\leq\), we
obtain \((S_2,H_2) < (S_2',H_2')\).
In view of the first case considered in this proof,
this implies \(S_2 = H_2'\) and \(H_2 = \emptyset\). Thus, by the definition
of a set pair, we have \(S_2 \cap S_2' = \emptyset\).
Moreover, \((S_2,H_2) < (S_1,H_1) \leq (S_2',H_2') \)
simplifies to \((S_2,\emptyset) < (S_1,H_1) \leq (S_2',S_2)\).
In view of the definition of \(\leq\), the latter can only hold if
\(S_2 = H_1 \neq \emptyset\). By (PL3), this implies
\((H_1,\emptyset) = (S_2,H_2) \in \mathcal{S}_1\), as required. 
\end{proof}

\begin{lemma}
\label{lem:technical:rank:function:graded:poset}
Let \(\mathcal{S}_1,\mathcal{S}_2 \in \mathfrak{P}(X)\) with
\(\mathcal{S}_1 \prec \mathcal{S}_2\). Then
\[1 \leq \lvert \mathcal{P}(\mathcal{S}_2) \rvert - \lvert \mathcal{H}(\mathcal{S}_2) \rvert < \lvert \mathcal{P}(\mathcal{S}_1) \rvert - \vert \mathcal{H}(\mathcal{S}_1) \rvert \leq \lvert X \rvert.\]
If \((\lvert \mathcal{P}(\mathcal{S}_1) \rvert - \lvert \mathcal{H}(\mathcal{S}_1)\rvert) - (\lvert \mathcal{P}(\mathcal{S}_2) \rvert - \lvert \mathcal{H}(\mathcal{S}_2) \rvert) \geq 2\)
then there exists \(\mathcal{S}_3 \in \mathfrak{P}(X)\) with
\(\mathcal{S}_1 \prec \mathcal{S}_3 \prec \mathcal{S}_2\).
\end{lemma}

\begin{proof}
In view of (PL1), we have
\(1 \leq \lvert \mathcal{P}(\mathcal{S}) \rvert \leq \lvert X \rvert\)
for all \(\mathcal{S} \in \mathfrak{P}(X)\). Moreover, in view of (PL3), 
we have $\lvert \mathcal{P}(\mathcal{S}) \rvert \geq 3 \lvert \mathcal{H}(\mathcal{S}) \rvert$.
This implies
\(1 \leq \lvert \mathcal{P}(\mathcal{S}) \rvert - \lvert \mathcal{H}(\mathcal{S}) \rvert \leq \lvert X \rvert\).

Next consider \(\mathcal{S}_1,\mathcal{S}_2 \in \mathfrak{P}(X)\) with
\(\mathcal{S}_1 \prec \mathcal{S}_2\). We first show that, for all
\(S' \in \mathcal{P}(\mathcal{S}_1)\), there exists a unique
\(S'' \in \mathcal{P}(\mathcal{S}_2)\) with \(S' \subseteq S''\).
In view of (PL2), there exists a unique set pair \((S_1,H_1) \in \mathcal{S}_1\)
with \(S' = S_1\) and, in view of \(\mathcal{S}_1 \prec \mathcal{S}_2\), there must exist
a set pair \((S_2,H_2) \in \mathcal{S}_2\) with \((S_1,H_1) \leq (S_2,H_2)\).
Therefore, by the definition of~\(\leq\), one of the following must hold:
\begin{itemize}
\item
\(S_1 \cup H_1 \subseteq S_2\). Then we put \(S''=S_2\).
\item
\(S_1 \cup H_1 \subseteq H_2\). This implies \(H_2 \neq \emptyset\) and thus,
by (PL3), \(H_2 \in \mathcal{P}(\mathcal{S}_2)\). We put \(S'' = H_2\).
\item
\(S_1 \subsetneq S_2\) and \(H_1 = H_2 \neq \emptyset\). Then we put \(S'' = S_2\).
\end{itemize}
In each case, we have \(S' \subseteq S''\) for some 
\(S'' \in \mathcal{P}(\mathcal{S}_2)\) and, in view of~(PL1), \(S''\) is
unique, as claimed. This implies that we obtain a map
\(q : \mathcal{S}_1 \rightarrow \mathcal{S}_2\) by assigning to
each \((S',H') \in \mathcal{S}_1\) the unique \((S'',H'') \in \mathcal{S}_2\)
with \(S' \subseteq S''\).
In particular, we have
$\lvert \mathcal{P}(\mathcal{S}_1)\rvert \geq \lvert \mathcal{P}(\mathcal{S}_2) \rvert$.

To establish $\lvert \mathcal{P}(\mathcal{S}_2) \rvert - \lvert \mathcal{H}(\mathcal{S}_2) \rvert <
\lvert \mathcal{P}(\mathcal{S}_1) \rvert - \lvert \mathcal{H}(\mathcal{S}_1)\rvert$, 
put \(k = \lvert \mathcal{P}(\mathcal{S}_1)\rvert - \rvert \mathcal{P}(\mathcal{S}_2)\rvert\).
Let~\(\ell_1\) denote the number of \(H' \in \mathcal{H}(\mathcal{S}_1)\)
with \(H' \not \in \mathcal{H}(\mathcal{S}_2)\).
Note that, in view of (PL3), for each such \(H'\), there exist precisely
two set pairs \((S_1',H_1'),(S_2',H_2') \in \mathcal{S}_1\)  with
\(H_1' = H_2' = H'\) and, in view of \(\mathcal{S}_1 \prec \mathcal{S}_2\),
there must exist some \(S'' \in \mathcal{P}(\mathcal{S}_2)\) with
\(S_1' \cup S_2' \cup H' \subseteq S''\). This implies \(k \geq 2\ell_1\).
Thus, letting \(\ell_2\) denote the number of \(H'' \in \mathcal{H}(\mathcal{S}_2)\)
with \(H'' \not \in \mathcal{H}(\mathcal{S}_1)\), we have
\begin{align*}
\lvert\mathcal{P}(\mathcal{S}_2)\rvert - \lvert \mathcal{H}(\mathcal{S}_2) \rvert
&= \lvert \mathcal{P}(\mathcal{S}_2) \rvert - \lvert \mathcal{H}(\mathcal{S}_1) \rvert + \ell_1 - \ell_2\\
&= \lvert \mathcal{P}(\mathcal{S}_1) \rvert -  \lvert \mathcal{H}(\mathcal{S}_1) \rvert + \ell_1 - \ell_2 - k\\
&\leq \lvert \mathcal{P}(\mathcal{S}_1) \rvert -  \lvert \mathcal{H}(\mathcal{S}_1) \rvert - \ell_1 - \ell_2.
\end{align*}
Thus, if \(\ell_1 + \ell_2 > 0\) we immediately have 
$\lvert \mathcal{P}(\mathcal{S}_2) \rvert - \lvert \mathcal{H}(\mathcal{S}_2) \rvert
< \lvert \mathcal{P}(\mathcal{S}_1) \rvert - \lvert \mathcal{H}(\mathcal{S}_1) \rvert$.
If \(\ell_1 + \ell_2 = 0\) we have
\(\mathcal{H}(\mathcal{S}_1) = \mathcal{H}(\mathcal{S}_2)\).
This implies, in view of \(\mathcal{S}_1 \prec \mathcal{S}_2\),
that we cannot have \(\mathcal{P}(\mathcal{S}_1) = \mathcal{P}(\mathcal{S}_2)\),
that is, we must have \(k > 0\) and, thus, we also obtain
$\lvert \mathcal{P}(\mathcal{S}_2)\rvert - \lvert \mathcal{H}(\mathcal{S}_2)\rvert
< \lvert\mathcal{P}(\mathcal{S}_1)\rvert - \lvert\mathcal{H}(\mathcal{S}_1)\rvert$, as required.

Now assume that \((\lvert\mathcal{P}(\mathcal{S}_1)\rvert - \lvert\mathcal{H}(\mathcal{S}_1)\rvert) - (\lvert\mathcal{P}(\mathcal{S}_2)\rvert - \lvert\mathcal{H}(\mathcal{S}_2)\rvert) \geq 2\).
First consider the case that there exist two distinct
\((S_1'',H_1''),(S_2'',H_2'') \in \mathcal{S}_2\) with
$\lvert q^{-1}(S_i'',H_i'')\rvert \geq 2$,
\(i \in \{1,2\}\). Then we put
\[\mathcal{S}_3 = (\mathcal{S}_1 - q^{-1}(S_1'',H_1'')) \cup \{(S_1'',H_1'')\}.\]
Next consider the case that there exists \((S'',H'') \in \mathcal{S}_2\)
with $\lvert q^{-1}(S'',H'')\rvert \geq 3$ and \(H' = \emptyset\)
for all \((S',H') \in q^{-1}(S'',H'')\). Then we select two distinct
\((S_1',\emptyset),(S_2',\emptyset) \in q^{-1}(S'',H'')\) and put
\[\mathcal{S}_3 = (\mathcal{S}_1 - \{(S_1',\emptyset),(S_2',\emptyset)\})
\cup \{(S_1' \cup S_2',\emptyset)\}.\]
The remaining case to consider is that there 
exists \((S'',H'') \in \mathcal{S}_2\) such that $\lvert q^{-1}(S'',H'')\rvert \geq 4$
and there are three distinct \((S_1',H_1'),(S_2',H_2'),(S_3',H_3')  \in q^{-1}(S'',H'')\)
with \(H_1' = \emptyset\) and \(H_2'=H_3'=S_1'\). Then we put
\[\mathcal{S}_3 = (\mathcal{S}_1 - \{(S_1',H_1'),(S_2',H_2'),(S_3',H_3')\})
\cup \{(S_1' \cup S_2' \cup S_3',\emptyset)\}.\]
In each case, by construction, we immediately have
\(\mathcal{S}_1 \prec \mathcal{S}_3 \prec \mathcal{S}_2\).
\end{proof}

\section{The poset $(\mathfrak{P}(X),\preceq)$}
\label{sec:poset:p:x}

In this section, we prove that~\(\preceq\)
is a partial ordering on~\(\mathfrak{P}(X)\). We also 
give a formula for counting the number of elements in
the resulting poset  $(\mathfrak{P}(X),\preceq)$.

We first recall some standard poset concepts (see e.g.~\cite{T95a}).
A (finite) \emph{poset} \((M,R)\) consists of a finite non-empty
set \(M\) and a binary relation~\(R \subseteq M \times M\) on~\(M\) that is reflexive,
transitive and antisymmetric. 
An element \(m \in M\) is \emph{minimum} (\emph{maximum}) if
\((m,a) \in R\) (\((a,m) \in R\)) holds for all \(a \in M\).
A poset is \emph{bounded} if it has a minimum and a maximum element
and these elements are then necessarily unique.
Two elements \(a,b \in M\) are
\emph{comparable} if \((a,b) \in R\) or \((b,a) \in R\).
A \emph{chain}~\(C\) is a non-empty subset of~\(M\) of pairwise
comparable elements. The \emph{length} of a chain~\(C\) is $\lvert C \rvert - 1$.
A chain is \emph{maximal} if it is not contained in some strictly longer chain. 
A poset is \emph{graded} if every maximal chain has the same length.
The \emph{height function}\footnote{Usually called \emph{rank function}
of the graded poset. We use height function instead to avoid confusion with
the rankings of rooted \(X\)-cactuses.}
\(h\) of a graded poset \((M,R)\) assigns to every
element \(a \in M\) the length \(h(a)\) of a longest chain \(C\)
with \((b,a) \in R\) for all \(b \in C\).  

\begin{proposition}
\label{prop:poset:p:x}
\((\mathfrak{P}(X),\preceq)\) is a bounded graded poset
with minimum element \(\{(\{x\},\emptyset) : x \in X\}\) and
maximum element \(\{(X,\emptyset)\}\). The height function
of this poset is \(h : \mathfrak{P}(X) \rightarrow \{0,1,\dots,\lvert X \rvert-1\}\)
with \(h(\mathcal{S}) = \lvert X \rvert - \lvert \mathcal{P}(\mathcal{S})\rvert + \lvert \mathcal{H}(\mathcal{S})\rvert\).
\end{proposition}

\begin{proof}
We first show that \((\mathfrak{P}(X),\preceq)\) is a poset.
It follows immediately from the definition of the binary relation~\(\preceq\)
that it is reflexive. Moreover, in view of
Lemma~\ref{lem:technical:rank:function:graded:poset},
we cannot have two distinct \(\mathcal{S}_1,\mathcal{S}_2 \in \mathfrak{P}(X)\)
with \(\mathcal{S}_1 \preceq \mathcal{S}_2\) and \(\mathcal{S}_2 \preceq \mathcal{S}_1\),
implying that~\(\preceq\) is also antisymmetric.

It remains to show that \(\preceq\) is transitive. 
Consider set pair systems \(\mathcal{S}_1\), \(\mathcal{S}_2\)
and \(\mathcal{S}_3\) with 
\(\mathcal{S}_1 \preceq \mathcal{S}_2 \preceq \mathcal{S}_3\).
Then, in view of (SP1),
for all \((S_1,H_1) \in \mathcal{S}_1\), there exists
some \((S_2,H_2) \in \mathcal{S}_2\) with \((S_1,H_1) \leq (S_2,H_2)\)
and, again in view of (SP1), there also exists some
\((S_3,H_3) \in \mathcal{S}_3\) with \((S_2,H_2) \leq (S_3,H_3)\).
By the transitivity of \(\leq\), we obtain
\((S_1,H_1) \leq (S_3,H_3)\), as required.

Next consider some \((S_3,H_3) \in \mathcal{S}_3\) with \(H_3 \neq \emptyset\).
First assume that there exists some \((S_2,H_2) \in \mathcal{S}_2\)
with \(H_2=H_3\). Then, by (SP2), there also exists
\((S_2,H_2) \in \mathcal{S}_2\) with \(H_2=H_3\) and
\((S_2,H_2) \leq (S_3,H_3)\). Now, if there exists some
\((S_1,H_1) \in \mathcal{S}_1\) with \(H_1 = H_2 = H_3\),
then, by (SP2), there also exists such a set pair in \(\mathcal{S}_1\)
with \((S_1,H_1) \leq (S_2,H_2) \leq (S_3,H_3)\).
Hence, by the transitivity of \(\leq\), we have
\((S_1,H_1) \leq (S_3,H_3)\), as required.

Next assume that there exists no \((S_2,H_2) \in \mathcal{S}_2\)
with \(H_2=H_3\). It suffices to show that this implies that
there exists no \((S_1,H_1) \in \mathcal{S}_1\)
with \(H_1=H_3\). So, assume for a contradiction that
there exists some \((S_1,H_1) \in \mathcal{S}_1\)
with \(H_1=H_3\neq \emptyset\). Put \(H=H_1\).
In view of \(\mathcal{S}_1 \preceq \mathcal{S}_2\),
there must exist some \((S_2,H_2) \in \mathcal{S}_2\)
with \((S_1,H) \leq (S_2,H_2)\). Note that \(H_2 \neq H\) combined with
the definition of \(\leq\) implies
\(S_1 \cup H \subseteq S_2\) or \(S_1 \cup H \subseteq H_2\).
Moreover, in view of \(\mathcal{S}_2 \preceq \mathcal{S}_3\),
there must exist some \((S_3',H_3') \in \mathcal{S}_3\)
with \((S_2,H_2) \leq (S_3',H_3')\). This implies
that \(S_1 \cup H \subseteq S_3'\) or \(S_1 \cup H \subseteq H_3'\).
But then, \(H \subsetneq S_3'\) or \(H \subsetneq H_3'\) must
hold in contradiction to (PL1). 
Thus \(\mathcal{S}_1 \preceq \mathcal{S}_3\) holds,
establishing that \(\preceq\) is transitive and, thus,
\((\mathfrak{P}(X),\preceq)\) is a poset.

Next we show that \(\{(\{x\},\emptyset) : x \in X\}\)
and \(\{(X,\emptyset)\}\) are the minimum and maximum element,
respectively, in \((\mathfrak{P}(X),\preceq)\). Clearly,
\(\{(\{x\},\emptyset) : x \in X\}\) and \(\{(X,\emptyset)\}\) are
both polestar systems and, thus, elements of \(\mathfrak{P}(X)\).
Consider any \(\mathcal{S} \in \mathfrak{P}(X)\). Then, for all
\((S,H) \in \mathcal{S}\), we have \(S \cup H \subseteq X\),
implying \((S,H) \leq (X,\emptyset)\) and, thus,
\(\mathcal{S} \preceq \{(X,\emptyset)\}\).
Similarly, in view of (PL1), for all \(x \in X\), there must exist
some \((S,H) \in \mathcal{S}\) with \(x \in S\), implying that
\((\{x\},\emptyset) \leq (S,H)\). Thus,
\(\{(\{x\},\emptyset) : x \in X\} \preceq \mathcal{S}\).
It follows that \((\mathfrak{P}(X),\preceq)\) is a bounded poset.

That \((\mathfrak{P}(X),\preceq)\) is a graded poset with height
function~\(h\) is now an immediate consequence of
Lemma~\ref{lem:technical:rank:function:graded:poset} in view of
\(h(\{(\{x\},\emptyset) : x \in X\}) = 0\) and
\(h(\{(X,\emptyset)\}) = \lvert X \rvert -1\). 
\end{proof}

The next corollary describes the relationship between~\((\mathfrak{P}(X),\preceq)\)
and the poset \((\mathfrak{B}(X),\sqsubseteq)\) of partitions of~\(X\).
Two posets \((M_1,R_1)\) and \((M_2,R_2)\) are \emph{isomorphic} if there
exists a bijective map \(f: M_1 \rightarrow M_2\) such that,
for all \(a,b \in M_1\), \((a,b) \in R_1\) if and only if \((f(a),f(b)) \in R_2\).

\begin{corollary}
\label{cor:poset:partitions}
The restriction of the poset~\((\mathfrak{P}(X),\preceq)\) to those
\(\mathcal{S} \in \mathfrak{P}(X)\) with \(\mathcal{H}(\mathcal{S}) = \emptyset\)
is isomorphic to the poset \((\mathfrak{B}(X),\sqsubseteq)\) of partitions of~\(X\).
\end{corollary}

\begin{proof}
We map any \(\mathcal{S} \in \mathfrak{P}(X)\) with
\(\mathcal{H}(\mathcal{S}) = \emptyset\) to
the partition \(\mathcal{P}(\mathcal{S}) \in \mathfrak{B}(X)\). This
map is bijective.
Moreover, for \(\mathcal{S}_1,\mathcal{S}_2 \in \mathfrak{P}(X)\)
with \(\mathcal{H}(\mathcal{S}_1) = \mathcal{H}(\mathcal{S}_2) = \emptyset\) we have
\(\mathcal{S}_1 \preceq \mathcal{S}_2\) if and only if
for all \(A_1 \in \mathcal{P}(\mathcal{S}_1)\) there exists
some \(A_2 \in \mathcal{P}(\mathcal{S}_2)\) with \(A_1 \subseteq A_2\),
as required.
\end{proof}

In the remaining part of this section, we give a formula for the
number~\(\lambda_n = \lvert\mathfrak{P}(X)\rvert\) of polestar
systems on a set \(X\) with \(n \geq 1\)
elements. The values of \(\lambda_n\) for \(n = 1,2,\dots,8\) are
1, 2, 8, 45, 277, 1853, 14065, 122118. 
For \(k \in \{1,2,\dots,n\}\), we denote by \(\alpha_{n,k}\)
the \emph{Stirling number} of the second kind,
that is, the number of partitions
of~\(X\) into~\(k\) subsets. In addition, for
\(\ell \in \{0,1,\dots,\lfloor \frac{k}{3} \rfloor\}\), we denote by \(\beta_{k,\ell}\)
the number of partitions of a set with~\(k\) elements into \(\ell\) subsets
with three elements and \(k-3\ell\) subsets with one element.
It is known~\cite{OEISa} that
\[\beta_{k,\ell} = \frac{k!}{6^{\ell} \cdot \ell! \cdot (k-3\ell)!}.\]

\begin{proposition}
\label{prop:size:px}
For all \(n \geq 1\) we have
\begin{equation}
\label{eq:number:polestar}
\lambda_n = \sum_{k=1}^n \alpha_{n,k} \cdot \left (\sum_{\ell=0}^{\lfloor \frac{k}{3} \rfloor}
\beta_{k,\ell} \cdot 3^{\ell} \right ).
\end{equation}
\end{proposition}

\begin{proof}
Let \(X\) be a set with \(n \geq 1\) elements.
Consider \(\mathcal{S} \in \mathfrak{P}(X)\) and put \(k =\lvert\mathcal{P}(\mathcal{S})\rvert\).
By the definition of a polestar system,
\(\mathcal{S}\) arises from \(\mathcal{P}(\mathcal{S})\) by forming, for some
\(\ell \in \{0,1,\dots,\lfloor \frac{k}{3} \rfloor\}\),
a partition \(\Pi(\mathcal{P}(\mathcal{S}))\)
of \(\mathcal{P}(\mathcal{S})\) into \(\ell\) subsets
with three elements and \(k-3\ell\) subsets with one element.
Each 1-element set \(\{S\} \in \Pi(\mathcal{P}(\mathcal{S}))\)
yields the set pair \((S,\emptyset)\).
For each 3-element set \(\{S_1,S_2,S_3\} \in \Pi(\mathcal{P}(\mathcal{S}))\) we select
\(i \in \{1,2,3\}\) and obtain the three set pairs \((S_i,\emptyset)\),
\((S_j,S_i)\), \(j \in \{1,2,3\} - \{i\}\).

Formula~(\ref{eq:number:polestar}) directly reflects the process described
above for obtaining a polestar system from a fixed partition
of \(X\) into \(k\) subsets. In view of the fact that
every partition of~\(X\) yields a different
collection of polestar systems on~\(X\),
we form the outer sum over the values of~\(k\).
The inner sum then accounts for the number of polestar
systems that arise from any fixed partition of \(X\) into~\(k\) subsets.
\end{proof}

\section{Encoding ranked $X$-cactuses}
\label{sec:encoding:ranked:x:network}

In this section, we show in~Theorem~\ref{theo:chains:networks:1:to:1}
that we can encode (isomorphism classes)
of ranked \(X\)-cactuses in terms of the chains
in the poset \((\mathfrak{P}(X),\preceq)\).
We begin by giving a precise statement of this result.
We call two equidistant \(X\)-cactuses 
\((\mathcal{N}'=((V',A'),\varphi'),t')\) 
and \((\mathcal{N}''=((V'',A''),\varphi''),t'')\) \emph{isomorphic}
if there exists a DAG-isomorphism \(f:V' \rightarrow V''\) such that 
\begin{itemize}
\setlength{\itemindent}{15pt}
\item[(IC1)]
\(f(\varphi'(x)) = \varphi''(x)\) for all \(x \in X\) and
\item[(IC2)]
\(t'(v) = t''(f(v))\) for all \(v \in V'\).
\end{itemize}
Note that this definition includes isomorphisms between ranked \(X\)-cactuses
as a special case. For rooted \(X\)-cactuses without a time-stamp function
to be isomorphic, condition~(IC2) is not required. We
now state the aforementioned result.

\begin{theorem}
\label{theo:chains:networks:1:to:1}
There is a one-to-one correspondence between chains in the poset
\((\mathfrak{P}(X),\preceq)\) that contain the maximum element
\(\{(X,\emptyset)\}\) and (isomorphism classes of) ranked \(X\)-cactuses.
The length of the chain equals the size of the ranking of the 
corresponding ranked \(X\)-cactus. Maximal chains correspond
to binary ranked \(X\)-cactuses with rankings of size $\lvert X \rvert-1$.
\end{theorem}

To prove this theorem, note that 
by Lemmas~\ref{lem:ranking:equals:partial:ordering}
and~\ref{lem:ranked:networks:yield:partition:like}, every
ranked \(X\)-cactus corresponds to a chain~\(\mathfrak{C}\)
in \((\mathfrak{P}(X),\preceq)\) with \(\{(X,\emptyset)\} \in \mathfrak{C}\).
Moreover, by~Lemma~\ref{lem:size:ranking},
we have $\lvert\mathfrak{C}\rvert \leq \lvert X \rvert - 1$
for such a chain with equality holding if and only if the ranked \(X\)-cactus is binary. 
Thus, to prove Theorem~\ref{theo:chains:networks:1:to:1}, it suffices to show that
for all chains \(\mathfrak{C}\) in \((\mathfrak{P}(X),\preceq)\) with
\(\{(X,\emptyset)\} \in \mathfrak{C}\) there exists, up to isomorphism,
a unique ranked \(X\)-cactus~\((\mathcal{N},r)\)
with \(\mathfrak{C} = \{\mathcal{S}_i(\mathcal{N}) : 0 \leq i \leq \sigma(r)\}\).
This follows immediately from 
Lemmas~\ref{lem:set:pairs:compressed:network} and~\ref{lem:modify:and:rank}
below, and will be done in two steps.
First, for any chain \(\mathfrak{C} \subseteq \mathfrak{P}(X)\)
with \(\{(X,\emptyset)\} \in \mathfrak{C}\), we form the set pair system
\(\mathcal{S}(\mathfrak{C}) = \bigcup_{\mathcal{S}' \in \mathfrak{C}} \mathcal{S}'\)
consisting of all set pairs that occur in the polestar
systems in~\(\mathfrak{C}\) and construct a suitable
rooted, compressed, phylogenetic \(X\)-cactus
\(\mathcal{N}(\mathfrak{C})\) (see Lemma~\ref{lem:set:pairs:compressed:network}).
Second, we perform some technical modifications on \(\mathcal{N}(\mathfrak{C})\),
if necessary, to obtain \(\mathcal{N}\) and then construct a suitable ranking~\(r\)
(see Lemma~\ref{lem:modify:and:rank}).

\begin{lemma}
\label{lem:set:pairs:compressed:network}
For all chains \(\mathfrak{C}\) in \((\mathfrak{P}(X),\preceq)\) with
\(\{(X,\emptyset)\} \in \mathfrak{C}\) there exists, up to isomorphism,
a unique rooted, compressed, phylogenetic \(X\)-cactus
\(\mathcal{N}(\mathfrak{C})\) with
\(\mathcal{S}(\mathcal{N}(\mathfrak{C})) = \mathcal{S}(\mathfrak{C}) \cup \{(\{x\},\emptyset) : x \in X\}\).
\end{lemma}

\begin{proof}
Put \(\mathcal{S} = \mathcal{S}(\mathfrak{C}) \cup \{(\{x\},\emptyset) : x \in X\}\).
We show below that \(\mathcal{S}\) satisfies certain
properties~(NC1)-(NC5). We do this to then apply~\cite[Theorem~5]{HMS21a},
which states that if a set pair system \(\mathcal{S}'\) on \(X\)
has these properties there exists, up to isomorphism, a unique
rooted, compressed, phylogenetic
\(X\)-cactus \(\mathcal{N}(\mathcal{S}')\)
with \(\mathcal{S}' = \mathcal{S}(\mathcal{N}(\mathcal{S}'))\), as required.
In the following we first state each of the properties~(NC1)-(NC5) and then
verify that \(\mathcal{S}\) has this property.

(NC1) -- \((X,\emptyset) \in \mathcal{S}\):\\
This is clearly the case.

(NC2) -- \((\{x\},\emptyset) \in \mathcal{S}\), for all \(x \in X\):\\
By construction of \(\mathcal{S}\), this is the case.

(NC3) -- For every \((S,H) \in \mathcal{S}\)
with \(H \neq \emptyset\), we have \((H,\emptyset) \in \mathcal{S}\):\\
Consider any \((S,H) \in \mathcal{S}\) with \(H \neq \emptyset\).
Then, by construction, there must exist some \(\mathcal{S}' \in \mathfrak{C}\)
with \((S,H) \in \mathcal{S}'\). In view of (PL3)
we must have \((H,\emptyset) \in \mathcal{S}'\). Thus, by the definition of
\(\mathcal{S}\), it follows that 
\((H,\emptyset) \in \mathcal{S}\), as required. 

(NC4) -- For any two distinct
\((S_1,H_1), (S_2,H_2) \in \mathcal{S}\)
one of (i)~\((S_1,H_1) < (S_2,H_2)\),
(ii)~\((S_2,H_2) < (S_1,H_1)\),
(iii)~\((S_1 \cup H_1) \cap (S_2 \cup H_2) = \emptyset\), or
(iv)~\(S_1 \cap S_2 = \emptyset\) and \(H_1 = H_2 \neq \emptyset\)
holds:\\
Consider \((S_1,H_1), (S_2,H_2) \in \mathcal{S}\)
with \((S_1,H_1) \neq (S_2,H_2)\).
By construction, there must exist \(\mathcal{S}_1,\mathcal{S}_2 \in \mathfrak{C}\)
with \((S_1,H_1) \in \mathcal{S}_1\) and \((S_2,H_2) \in \mathcal{S}_2\). Without loss
of generality we may assume that \(\mathcal{S}_1 \preceq \mathcal{S}_2\). 

First we consider the case~\(\mathcal{S}_1 = \mathcal{S}_2\). Then, in view of
(PL1) and (PL2), we have \(S_1 \cap S_2 = \emptyset\). Thus, if
\(H_1 = H_2 \neq \emptyset\), we are done. Otherwise, in view of (PL3) and (PL1),
we must have \(H_1 \cap H_2 = \emptyset\) and, thus,
\((S_1 \cup H_1) \cap (S_2 \cup H_2) = \emptyset\), as required.

Next consider the case~\(\mathcal{S}_1 \prec \mathcal{S}_2\). 
Then there must exist some \((S,H) \in \mathcal{S}_2\)
with \((S_1,H_1) \leq (S,H)\). If \((S,H) = (S_2,H_2)\)
we immediately have \((S_1,H_1) \leq (S_2,H_2)\) and are done.
So assume \((S,H) \neq (S_2,H_2)\). In view (PL2),
this implies \(S \cap S_2 = \emptyset\). Thus, by the definition of~\(\leq\)
one of the following must hold:
\begin{itemize}
\item
\(S_1 \cup H_1 \subseteq S\):
Then, by the definition of set pairs,
\((S_1 \cup H_1) \cap H = \emptyset\) and,
in view of \(S \cap S_2 = \emptyset\),
also \((S_1 \cup H_1) \cap S_2 = \emptyset\).
Thus, if \(S \cap H_2 = \emptyset\)
we have \((S_1 \cup H_1) \cap (S_2 \cup H_2) = \emptyset\).
So, assume that \(S=H_2\).
Then we have \(S_1 \cup H_1 \subseteq H_2\) implying
\((S_1,H_1) < (S_2,H_2)\).
\item
\(S_1 \cup H_1 \subseteq H\):
Then, if \(H=H_2\) or \(H=S_2\),
we immediately have \((S_1,H_1) < (S_2,H_2)\).
Otherwise we must have
\(H \cap H_2 = \emptyset\) and \(H \cap S_2 = \emptyset\)
and, thus, \((S_1 \cup H_1) \cap (S_2 \cup H_2) = \emptyset\).
\item
\(S_1 \subsetneq S\) and \(H_1 = H \neq \emptyset\):
First note that this implies \(S \cap H_2 = \emptyset\)
because otherwise we would have \((S,\emptyset) \in \mathcal{S}_2\)
in view of (PL3), which is impossible in view of
\((S,H) \in \mathcal{S}_2\) and (PL2). Also note that if
\(H=S_2\) we must have \((S_2,H_2) = (H,\emptyset)\)
in view of (PL2), implying that \((S_1,H_1) < (S_2,H_2)\).
Finally, if \(H \cap S_2 = \emptyset\) we obtain
\((S_1 \cup H_1) \cap (S_2 \cup H_2) = \emptyset\).
\end{itemize}
This establishes that \(\mathcal{S}\) satisfies~(NC4).

(NC5) -- There are no three distinct 
\((S_1,H_1)\), \((S_2,H_2)\), \((S_3,H_3) \in \mathcal{S}\)
with \(H_1 = H_2 = H_3 \neq \emptyset\),
\(S_1 \cap S_2 = \emptyset\) and either \(S_1 \cup S_2 \subseteq S_3\) or
\((S_1 \cup S_2) \cap S_3 = \emptyset\):\\
Consider \(\mathcal{S}_1, \mathcal{S}_2, \mathcal{S}_3 \in \mathfrak{C}\)
with \(\mathcal{S}_1 \preceq \mathcal{S}_2 \preceq \mathcal{S}_3\).
Assume that there exist set pairs
\((S_1',H),(S_1'',H) \in \mathcal{S}_1\),
\((S_2',H),(S_2'',H) \in \mathcal{S}_2\) and
\((S_3',H),(S_3'',H) \in \mathcal{S}_3\)
with \(H \neq \emptyset\). By~(PL3),
there are precisely these two set pairs contained in
each of \(\mathcal{S}_1\), \(\mathcal{S}_2\) and \(\mathcal{S}_3\) for the fixed set~\(H\).
In view of (SP2), we may assume without loss of generality
that \((S_1',H) \leq (S_2',H) \leq (S_3',H)\)
and  \((S_1'',H) \leq (S_2'',H) \leq (S_3'',H)\),
implying that we have \(S_1' \subseteq S_2' \subseteq S_3'\)
and \(S_1'' \subseteq S_2'' \subseteq S_3''\).
But then it is impossible to select three distinct set pairs
\((S_1,H), (S_2,H), (S_3,H)\) from among
\[(S_1',H),(S_1'',H),(S_2',H),(S_2'',H),(S_3',H),(S_3'',H)\]
with \(S_1 \cap S_2 = \emptyset\) and either \(S_1 \cup S_2 \subseteq S_3\) or
\((S_1 \cup S_2) \cap S_3 = \emptyset\).
This establishes that \(\mathcal{S}\) satisfies~(NC5).
\end{proof}

Recall from Section~\ref{sec:preliminaries} that, for every
rooted \(X\)-cactus \(\mathcal{N}\), we denote by~\(\widehat{\mathcal{N}}^*\)
the associated rooted, compressed, phylogenetic \(X\)-cactus.

\begin{lemma}
\label{lem:modify:and:rank}
For all chains \(\mathfrak{C}\) in \((\mathfrak{P}(X),\preceq)\) with
\(\{(X,\emptyset)\} \in \mathfrak{C}\)
there exists, up to isomorphism, a unique ranked \(X\)-cactus
\((\mathcal{N},r)\) such that \(\widehat{\mathcal{N}}^* = \mathcal{N}(\mathfrak{C})\)
and \(\mathfrak{C} = \{\mathcal{S}_i(\mathcal{N}) : 0 \leq i \leq \sigma(r)\}\).
\end{lemma}

\begin{proof}
Consider the rooted, compressed, phylogenetic \(X\)-cactus
\(\mathcal{N}(\mathfrak{C}) = ((\widehat{V}^*,\widehat{A}^*),\widehat{\varphi}^*)\)
that exists by Lemma~\ref{lem:set:pairs:compressed:network}.
To obtain a suitable rooted \(X\)-cactus \(\mathcal{N} = ((V,A),\varphi)\)
with \(\mathcal{S}(\mathcal{N}) = \mathcal{S}(\mathfrak{C})\),
we take \(\mathcal{N}(\mathfrak{C})\) and modify it. 
The first modification applies to all \(x \in X\) with
\((\{x\},\emptyset) \not \in \mathcal{S}(\mathfrak{C})\)
and corresponds to reversing the addition of leaves
that was illustrated in Figure~\ref{fig:one:nested:x:networks}(b).
For each such \(x\), we contract the arc \((u,v) \in \widehat{A}^*\)
with \(v = \widehat{\varphi}^*(x)\) and put \(\varphi(x) = u\).
The second modification applies to all set pairs
\((S,\emptyset) \in \mathcal{S}(\mathfrak{C})\)
such that \((S,\emptyset) \in \mathcal{S}_1 \cap \mathcal{S}_2\) for
\(\mathcal{S}_1, \mathcal{S}_2 \in \mathfrak{C}\) with
\(\mathcal{S}_1 \prec \mathcal{S}_2\), \(S \not \in \mathcal{H}(\mathcal{S}_1)\)
and \(S \in \mathcal{H}(\mathcal{S}_2)\).
This implies, in view of the definition of the polestar
systems \(\mathcal{S}_i(\mathcal{N})\), \(0 \leq i \leq \lvert X \rvert-1\),
that we need to modify~\(\mathcal{N}(\mathfrak{C})\) to ensure that
\(\mathcal{N}\) contains two distinct vertices \(u\) and \(v\)
with \((S(u),H(u)) = (S(v),H(v)) = (S,\emptyset)\).
This corresponds to reversing the compression
that was illustrated in Figure~\ref{fig:one:nested:x:networks}(c).
Thus, in view of~(SH2), for each such set pair \((S,H)\), we locate the vertex
\(u \in \widehat{V}^*\) with \((S(u),H(u)) = (S,\emptyset)\)
and then expand the vertex \(u\) into an arc \((u,v)\)
such that the outgoing arcs of \(u\) become the outgoing arcs of \(v\)
and, for all \(x \in X\) with \(u=\widehat{\varphi}^*(x)\),
we put \(\varphi(x) = v\).

Note that the resulting rooted \(X\)-cactus~\(\mathcal{N}\)
need no longer be phylogenetic or compressed and that for all
ranked \(X\)-cactuses \((\mathcal{N}',r')\) with
\(\mathfrak{C} = \{\mathcal{S}_i(\mathcal{N}') : 0 \leq i \leq \sigma(r)'\}\)
we necessarily have that \(\mathcal{N}'\) is isomorphic to \(\mathcal{N}\)
in view of the fact that \(\widehat{\mathcal{N}'}^*\) and
\(\widehat{\mathcal{N}}^*\) must be isomorphic
by Lemma~\ref{lem:set:pairs:compressed:network}.

Thus, it remains to show that there exists
a unique ranking~\(r\) of the vertices
of \(\mathcal{N}\) to obtain a ranked \(X\)-cactus
\((\mathcal{N},r)\) with 
\(\mathfrak{C} = \{\mathcal{S}_i(\mathcal{N}) : 0 \leq i \leq \sigma(r)\}\).
Let \(c\) denote the length of \(\mathfrak{C}\)
and consider the sequence
\(\mathcal{S}_0 \prec \mathcal{S}_1 \prec \dots \prec \mathcal{S}_c\)
of the polestar systems in~\(\mathfrak{C}\).
The value \(r(u)\) for a vertex \(u \in V\)
that is not a reticulation vertex is defined by 
considering the set pair \((S(u),H(u))\) and putting
\(r(u)\) to be the smallest index \(0 \leq i \leq c\)
with \((S(u),H(u)) \in \mathcal{S}_i\).
Note that this is the only available choice for the rank of~\(u\).
The value \(r(u)\) of a reticulation vertex \(u\)
is defined to be equal to the rank of the parents of \(u\),
which, since \(\mathcal{N}\) is an rooted \(X\)-cactus,
cannot be reticulation vertices and have been assigned a rank already.

Next, we show that the map \(r: V \rightarrow \{0,1,\dots,c\}\)
defined above is a ranking of the vertices of~\(\mathcal{N}\).
First note that the value \(r(u)\) of a reticulation vertex \(u\) is
well-defined. Indeed, in view of~(SH3), we must have \(r(p_1) = r(p_2)\)
for the two parents~\(p_1\) and~\(p_2\) of~\(u\), that is,
the set pairs \((S(p_1),H(p_1))\) and \((S(p_2),H(p_2))\) 
with \(H(p_1) = H(p_2) = H\) are both
contained in the polestar system \(\mathcal{S}_i\)
with the smallest index~\(i\) such that
\(H \in \mathcal{H}(\mathcal{S}_i)\). This establishes~(TS3).

To establish~(TS1), consider any \(x \in X\).
By (PL1) there exists a unique set pair \((S,H) \in \mathcal{S}_0\) with \(x \in S\).
Then, by Lemma~\ref{lem:technical:ordering:consistent}, it suffices to consider the
following two cases:
\begin{itemize}
\item
There is precisely one \((S',H') \in \mathcal{S}(\mathfrak{C})\) with
\((S',H') < (S,H)\). Then we must have \((S',H') \in \mathcal{S}_0\),
\(H' = \emptyset\) and \(H=S'\). This implies that there exists a
reticulation vertex \(u\) in \(\mathcal{N}\) that is a leaf with
\((S(u),H(u)) = (S',H')\) and that \(u\) is the single child of
a vertex \(p\) with \((S(p),H(p)) = (S,H)\). Since \(x \in S\) and
\(S \cap S' = \emptyset\), we have \(\varphi(x) = p\). By construction,
we have \(r(p) = 0\), as required.
\item
There is no \((S',H') \in \mathcal{S}(\mathfrak{C})\) with
\((S',H') < (S,H)\). Then there exists a leaf \(u\) of \(\mathcal{N}\)
with \((S(u),H(u)) = (S,H)\) and we must have \(\varphi(x) = u\).
Again, by construction, we have \(r(u) = 0\), as required.
\end{itemize}

Now, we turn to~(TS2).
Consider an arc \((u,v)\) of \(\mathcal{N}\) such that \(v\) is not
a reticulation vertex. As mentioned in Section~\ref{sec:set:pair:systems},
since~\(v\) is a descendant of~\(u\), we have \((S(v),H(v)) \leq (S(u),H(u))\).
If \((S(v),H(v)) = (S(u),H(u))\) then, by the construction of \(\mathcal{N}\)
from \(\mathcal{N}(\mathfrak{C})\), \(u\) is a reticulation vertex
whose single child is \(v\) and there exist \(0 \leq i < j \leq c\)
with \(r(v) = i\) and \(r(u) = r(p_1) = r(p_2) = j\), where~\(p_1\)
and~\(p_2\) are the two parents of~\(u\). Similarly,
in view of Lemma~\ref{lem:technical:ordering:consistent}, if \((S(v),H(v)) < (S(u),H(u))\)
there also exist \(0 \leq i < j \leq c\) with \(r(v) = i\) and
\(r(u) = j\). This establishes~(TS2).

The last property required for the map \(r\) to be a ranking is that,
for all \(j \in \{0,1,\dots,c\}\), there exists a vertex \(u\) of \(\mathcal{N}\)
with \(r(u) = j\). (TS1)~implies that this is the case for \(j=0\).
So, consider \(j \geq 1\). Then, in view of
Lemma~\ref{lem:technical:rank:function:graded:poset}, there exists
some \((S,H) \in \mathcal{S}_j\) with \((S,H) \not \in \mathcal{S}_i\)
for all \(i < j\). Let \(u\) be a vertex of \(\mathcal{N}\) with
\((S(u),H(u)) = (S,H)\). If \(u\) is not a reticulation vertex,
we have \(r(u) = j\). If \(u\) is a reticulation vertex, 
we have \(r(u) = r(p_1) = r(p_2) = j\) for the two parents
\(p_1\) and \(p_2\) of \(u\) since, by (PL3), 
\((S(p_1),H(p_1))\) and \((S(p_2),H(p_2))\) are also both contained
in \(\mathcal{S}_j\) but not in \(\mathcal{S}_i\)
for all \(i < j\). 

To finish the proof of the lemma, we show that 
\(\mathcal{S}_j = \mathcal{S}_j(\mathcal{N})\) for all \(j \in \{0,1,\dots,c\}\).
We clearly have \(\mathcal{S}_c = \{(X,\emptyset)\} = \mathcal{S}_c(\mathcal{N})\).
Consider \(j < c\).
In view of Lemma~\ref{lem:ranked:networks:yield:partition:like},
(PL1) and (PL2) it suffices to show
that, for all \(u \in V_j\), we have \((S(u),H(u)) \in \mathcal{S}_j\).
Let \(p\) be the unique parent of \(u\). By the definition of \(V_j\)
given in Section~\ref{sec:set:pair:systems},
we have \(r(u) = i \leq j\) and \(r(p) = k > j\). 
In particular, we have \((S(u),H(u)) \in \mathcal{S}_i\)
and \((S(p),H(p)) \in \mathcal{S}_k\). 
In view of \(\mathcal{S}_i \preceq \mathcal{S}_j \prec \mathcal{S}_k\)
there must exist \((S',H') \in \mathcal{S}_j\) with
\((S(u),H(u)) \leq (S',H')\) and also some \((S'',H'') \in \mathcal{S}_k\) with
\((S',H') \leq (S'',H'')\). Since \(p\) is the parent
of \(u\) we have \((S(u),H(u)) \leq (S(p),H(p))\) and,
since all set pairs in \(\mathcal{S}(\mathfrak{C})\)
correspond to at least one vertex of \(\mathcal{N}\),
we must necessarily have \((S'',H'') = (S(p),H(p))\).
It follows that either \((S(u),H(u)) = (S',H') = (S(p),H(p))\)
or \((S(u),H(u)) = (S',H') < (S(p),H(p))\) holds,
implying \((S(u),H(u)) \in \mathcal{S}_j\), as required.
\end{proof}

As an immediate consequence of Theorem~\ref{theo:chains:networks:1:to:1}
we obtain Theorem~\ref{prop:pairwise:trees}, which we restate in the
following corollary using poset terminology.

\begin{corollary}
\label{cor:chains:rooted:x:trees}
There is a one-to-one correspondence between chains in the graded poset
\((\mathfrak{B}(X),\sqsubseteq)\) that contain~\(\{X\}\) and
isomorphism classes of ranked \(X\)-trees.
\end{corollary}

\begin{proof}
In view of the fact that a rooted \(X\)-cactus \(\mathcal{N}\)
is a rooted \(X\)-tree if and only if the associated set pair system
\(\mathcal{S}(\mathcal{N})\) does not contain a set pair \((S,H)\) with
\(H \neq \emptyset\), it follows by Theorem~\ref{theo:chains:networks:1:to:1}
that ranked \(X\)-trees correspond to chains \(\mathfrak{C}\) in the poset
\((\mathfrak{P}(X),\preceq)\) with \(\{(X,\emptyset)\} \in \mathfrak{C}\)
and \(\mathcal{H}(\mathcal{S})=\emptyset\) for all \(\mathcal{S} \in \mathfrak{C}\).
This implies, by Corollary~\ref{cor:poset:partitions},
that ranked \(X\)-trees correspond to chains in the
poset~\((\mathfrak{B}(X),\sqsubseteq)\) that contain the partition~\(\{X\}\).
\end{proof}

\section{The space of equidistant $X$-cactuses}
\label{sec:network:space}

\begin{figure}
\centering
\includegraphics[scale=0.9]{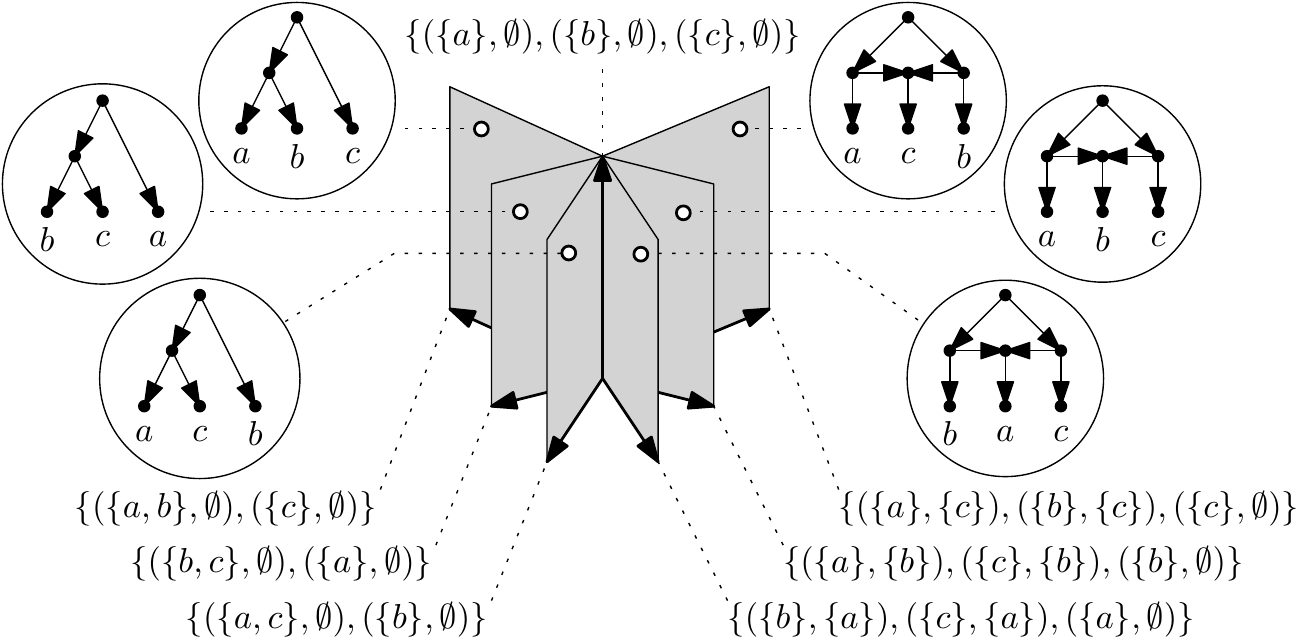}
\caption{The structure of \(\mathfrak{N}(X)\)
for \(X=\{a,b,c\}\).
The six 2-dimensional orthants are drawn shaded.
Each of these 2-dimensional orthants corresponds to an
isomorphism class of binary ranked \(X\)-cactuses.
Each axis corresponds to the indicated polestar
system on \(X\).}
\label{fig:orthants:networks:3:elements}
\end{figure}

We now define equidistant-cactus space, $\mathfrak{N}(X)$, and show
that it is a \cat(0)-metric space. 
The construction of $\mathfrak{N}(X)$  follows the outline 
presented at the start of Section~\ref{sec:encoding}.  More specifically, we 
put \(\mathfrak{P}^{\circ}(X) = \mathfrak{P}(X) - \{\{(X,\emptyset)\}\}\)
and let \(\mathcal{F}(\preceq)\) denote the set of chains in the subposet
\((\mathfrak{P}^{\circ}(X),\preceq)\) of the poset \((\mathfrak{P}(X),\preceq)\).
We then define $\mathfrak{N}(X)$ to be the orthant space of 
the order complex of \((\mathfrak{P}^{\circ}(X),\preceq)\).
Figure~\ref{fig:orthants:networks:3:elements} gives an example of
the structure of $\mathfrak{N}(X)$ for $X=\{a,b,c\}$. 

\begin{theorem}
\label{theo:metric:network:space}
\(\mathfrak{N}(X) = (\mathfrak{M}_{(\mathfrak{P}^{\circ}(X),\mathcal{F}(\preceq))},D_{(\mathfrak{P}^{\circ}(X),\mathcal{F}(\preceq))})\) is a \(\cat(0)\)-metric space
whose points are in one-to-one correspondence with
isomorphism classes of equidistant \(X\)-cactuses.
\end{theorem}

\begin{proof}
As an immediate consequence of the definition of a chain
as a set of pairwise comparable elements in a poset, we have
that \((\mathfrak{P}^{\circ}(X),\mathcal{F}(\preceq))\) is a flag complex
(cf. Section~\ref{sub:sec:orthant:spaces}).
Hence, \((\mathfrak{M}_{(\mathfrak{P}^{\circ}(X),\mathcal{F}(\preceq))},D_{(\mathfrak{P}^{\circ}(X),\mathcal{F}(\preceq))})\) is a \(\cat(0)\)-metric space.

It remains to show that the points of \(\mathfrak{M}_{(\mathfrak{P}^{\circ}(X),\mathcal{F}(\preceq))}\)
are in one-to-one correspondence with
isomorphism classes of equidistant \(X\)-cactuses.
Every \(\omega \in \mathfrak{M}_{(\mathfrak{P}^{\circ}(X),\mathcal{F}(\preceq))}\) corresponds,
up to isomorphism, to a unique equidistant \(X\)-cactus \((\mathcal{N},t)\) as follows.
Put \(\sigma = \lvert \supp(\omega) \rvert\) and
\(\mathfrak{C} = \supp(\omega) \cup \{\{(X,\emptyset)\}\}\).
Note that \(\mathfrak{C}\) is a chain in the poset \((\mathfrak{P}(X),\preceq)\).
Consider the sequence
\[\mathcal{S}_0 \prec \mathcal{S}_1 \prec \mathcal{S}_2 \prec
\dots \prec \mathcal{S}_{\sigma} = \{(X,\emptyset)\}\]
of the set pair systems in \(\mathfrak{C}\).
By~Theorem~\ref{theo:chains:networks:1:to:1}, there exists, up to isomorphism,
a unique ranked \(X\)-cactus \((\mathcal{N}=((V,A),\varphi),r)\)
with \(\sigma(r) = \sigma\) and \(\mathcal{S}_i = \mathcal{S}_i(\mathcal{N})\) for
all \(i \in \{0,1,2,\dots,\sigma\}\). The time-stamp function~\(t\) 
on the vertices of \(\mathcal{N}\) is then defined by putting
\[t(v) = \begin{cases}
0 &\text{if} \ r(v) = 0\\
\sum_{i=0}^{r(v)-1} \omega(\mathcal{S}_i) &\text{if} \ r(v) > 0
\end{cases}\]
for all \(v \in V\). Note that every
\(\omega' \in \mathfrak{M}_{(\mathfrak{P}^{\circ}(X),\mathcal{F}(\preceq))}\)
with \(\omega' \neq \omega\) and \(\supp(\omega') = \supp(\omega)\)
yields the same ranked \(X\)-cactus \((\mathcal{N},r)\) but a time-stamp
function~\(t' \neq t\) on the vertices of~\(\mathcal{N}\).
Also note that every equidistant \(X\)-cactus~\((\mathcal{N},t)\)
arises from some \(\omega \in \mathfrak{M}_{(\mathfrak{P}^{\circ}(X),\mathcal{F}(\preceq))}\)
as described above.
\end{proof}

To illustrate the proof of Theorem~\ref{theo:metric:network:space},
consider the equidistant \(X\)-cactus \((\mathcal{N},t)\) on \(X=\{a,b,c,d,e\}\) in
Figure~\ref{fig:equidistant:network}, which arises
from the point \(\omega \in \mathfrak{M}_{(\mathfrak{P}^{\circ}(X),\mathcal{F}(\preceq))}\) with
\(\supp(\omega) = \{\mathcal{S}_0,\mathcal{S}_1,\mathcal{S}_2,\mathcal{S}_3\}\),
where
\begin{align*}
\mathcal{S}_3 &= \{(\{a\},\{b\}),(\{b\},\emptyset),(\{c,d,e\},\{b\})\}\\
\mathcal{S}_2 &= \{(\{a\},\{b\}),(\{b\},\emptyset),(\{c\},\{b\}),(\{d,e\},\emptyset)\}\\
\mathcal{S}_1 &= \{(\{a\},\{b\}),(\{b\},\emptyset),(\{c\},\{b\}),(\{d\},\emptyset),(\{e\},\emptyset)\}\\
\mathcal{S}_0 &= \{(\{a\},\emptyset),(\{b\},\emptyset),(\{c\},\emptyset),(\{d\},\emptyset),(\{e\},\emptyset)\},
\end{align*}
and \(\omega(\mathcal{S}_0) = 0.8\), \(\omega(\mathcal{S}_1) = 0.4\),
\(\omega(\mathcal{S}_2) = 1.2\), \(\omega(\mathcal{S}_3) = 0.6\).

\begin{figure}
\centering
\includegraphics[scale=0.43]{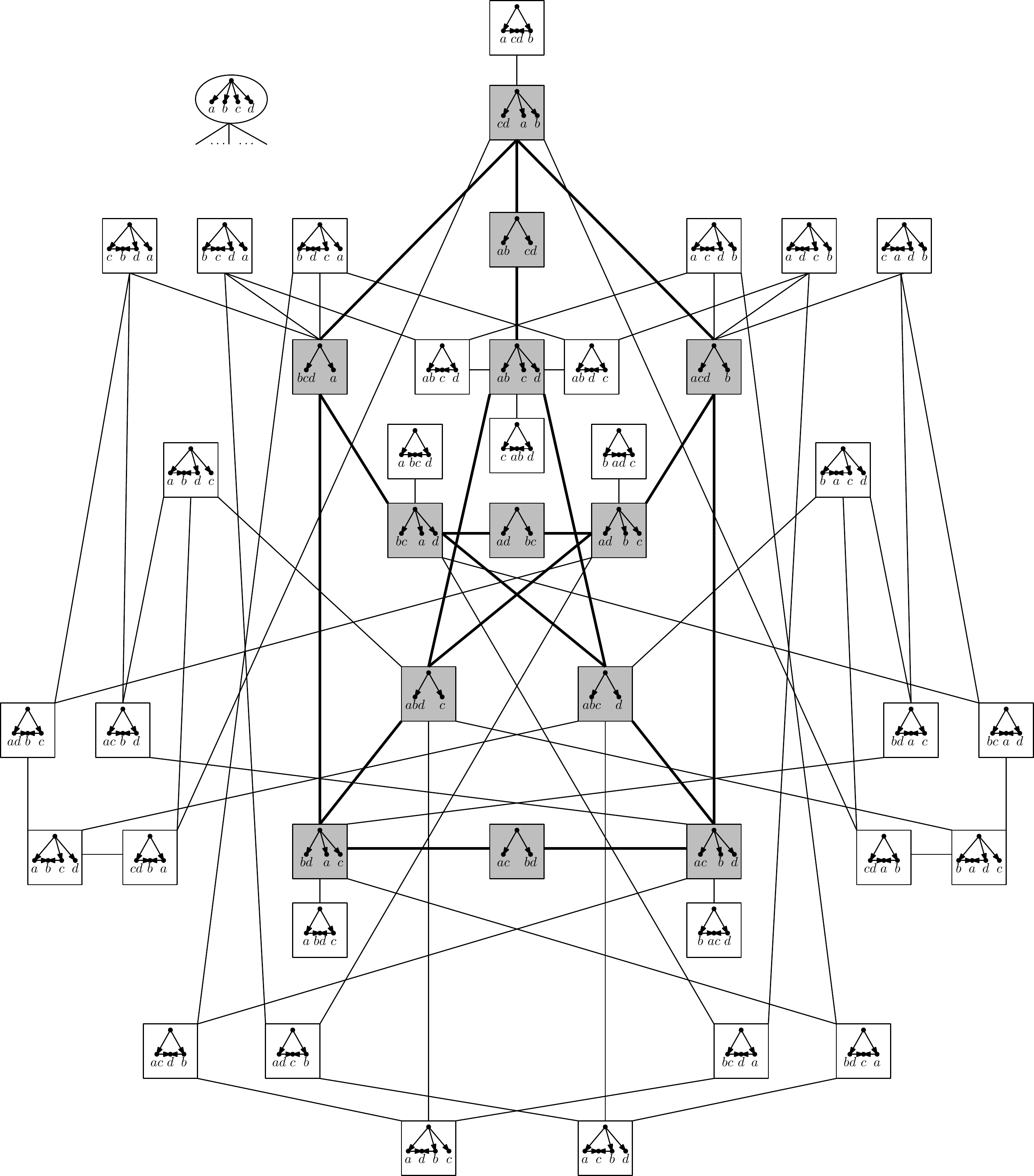}
\caption{The graph that determines the structure of the link 
of the origin \(\mathfrak{L}_{(\mathfrak{P}^{\circ}(X),\mathcal{F}(\preceq))}\) for \(X=\{a,b,c,d\}\).
The oval vertex is adjacent to all other vertices.
The ranked \(X\)-cactus displayed for each vertex
corresponds to the chain \(\{\mathcal{S},\{(X,\emptyset)\}\}\) for each
\(\mathcal{S} \in \mathfrak{P}^{\circ}(X)\).}
\label{figure:link:n:4}
\end{figure}

In general, as equidistant-cactus space is high-dimensional,
for $\lvert X \rvert \geq 4$ its structure is not easy to
visualize. However, to get some insights it can be useful 
to consider the so-called \emph{link of the origin}
\[\mathfrak{L}_{(\mathfrak{P}^{\circ}(X),\mathcal{F}(\preceq))} 
= \{\omega \in \mathfrak{M}_{(\mathfrak{P}^{\circ}(X),\mathcal{F}(\preceq))}
: \sum_{\mathcal{S} \in \mathfrak{P}^{\circ}(X)} \omega(\mathcal{S}) = 1\},\]
a geometric realization of the abstract simplicial
complex~\((\mathfrak{P}^{\circ}(X),\mathcal{F}(\preceq))\).
Since \((\mathfrak{P}^{\circ}(X),\mathcal{F}(\preceq))\) is a flag
complex, the structure of \(\mathfrak{L}_{(\mathfrak{P}^{\circ}(X),\mathcal{F}(\preceq))}\)
is completely determined
by the graph with vertex set \(\mathfrak{P}^{\circ}(X)\) in which two
distinct vertices are connected by an edge if and only if
they are comparable by~\(\preceq\). In 
Figure~\ref{figure:link:n:4} we present the link of the origin of $\mathfrak{N}(X)$  
for $\lvert X \rvert=4$. Note that, for this case, we have $\lvert\mathfrak{P}^{\circ}(X)\rvert = 44$
and that there are 14~vertices that correspond to rooted \(X\)-trees. The shaded vertices
in Figure~\ref{figure:link:n:4}
together with the oval vertex induce a subgraph that is isomorphic to the graph
corresponding to the link of the origin of $\tau$-space (i.e.
\(\mathfrak{M}_{(\mathfrak{B}^{\circ}(X),\mathcal{F}(\sqsubseteq))}\)),  which
is isomorphic to a subdivision of the Petersen graph 
(see also \cite[Fig.~3]{GD16}).

We conclude this section with a corollary of Theorem~\ref{theo:metric:network:space} 
that describes a relationship between $\tau$-space and equidistant-cactus space.

\begin{corollary}
\label{cor:tau:space}
The orthants of \(\mathfrak{M}_{(\mathfrak{B}^{\circ}(X),\mathcal{F}(\sqsubseteq))}\)
are in one-to-one correspondence
with the orthants \(\mathfrak{O}(A)\) of 
\(\mathfrak{M}_{(\mathfrak{P}^{\circ}(X),\mathcal{F}(\preceq))}\) for those
\(A \in \mathcal{F}(\preceq)\) with 
\(\mathcal{H}(\mathcal{S}) = \emptyset\) for all \(\mathcal{S} \in A\).
\end{corollary}

\begin{proof}
By definition, the orthants of \(\mathfrak{M}_{(\mathfrak{B}^{\circ}(X),\mathcal{F}(\sqsubseteq))}\)
are in one-to-one correspondence
with chains in \((\mathfrak{B}^{\circ}(X),\sqsubseteq)\).
By Corollary~\ref{cor:poset:partitions} and the definition of
\(\mathcal{F}(\preceq)\), such chains are in one-to-one correspondence
with chains \(\mathfrak{C}\) in \((\mathfrak{P}^{\circ}(X),\preceq)\)
for which \(\mathcal{H}(\mathcal{S}) = \emptyset\)
for all \(\mathcal{S} \in \mathfrak{C}\).
Again by definition, the latter chains are in one-to-one correspondence
with the orthants \(\mathfrak{O}(A)\) of
\(\mathfrak{M}_{(\mathfrak{P}^{\circ}(X),\mathcal{F}(\preceq))}\) for those
\(A \in \mathcal{F}(\preceq)\) with 
\(\mathcal{H}(\mathcal{S}) = \emptyset\) for all \(\mathcal{S} \in A\).
\end{proof}

We remark that the characterization of geodesic paths in
\cat(0)-orthant spaces in \cite[Corollary~6.19]{MOP15a}
holds for equidistant-cactus space~\(\mathfrak{N}(X)\).
This implies that, for any two points in \(\mathfrak{N}(X)\) that correspond 
to equidistant \(X\)-trees, all points on the unique geodesic path
between these two points also correspond to equidistant \(X\)-trees.
In other words,
\((\mathfrak{M}_{(\mathfrak{B}^{\circ}(X),\mathcal{F}(\sqsubseteq))},D_{(\mathfrak{P}^{\circ}(X),\mathcal{F}(\preceq))})\) is a convex subspace of
\(\mathfrak{N}(X) = (\mathfrak{M}_{(\mathfrak{P}^{\circ}(X),\mathcal{F}(\preceq))},D_{(\mathfrak{P}^{\circ}(X),\mathcal{F}(\preceq))})\).

\section{Conclusion}
\label{sec:conclusion}

We have introduced the space $\mathfrak{N}(X)$ of equidistant $X$-cactuses.
By deriving an encoding for ranked $X$-cactuses, we obtained
$\mathfrak{N}(X)$ as an orthant space and proved that it
is a \cat(0)-metric space. Thus, we can compute 
the distance in $\mathfrak{N}(X)$ between any two equidistant $X$-cactuses
and the unique geodesic path between them 
in polynomial time \cite{MOP15a}, compute approximations
of the Fr\'echet mean and variance as well as
of the median of a set of equidistant \(X\)-cactuses \cite{bacak2014computing,MOP15a},
and a central limit theorem holds \cite{barden2018logarithm}.  
There are several directions for future research and open questions including:

\begin{itemize}
\item
It would be interesting
to count the number~\(\nu_n\) of isomorphism classes 
of binary ranked \(X\)-cactuses with rankings of size~$\lvert X \rvert-1$.
In view of Theorem~\ref{theo:chains:networks:1:to:1}, this is
equivalent to counting the number of maximal chains in the graded
poset \((\mathfrak{P}(X),\preceq)\). Counting chains in certain types of
posets is a well-studied problem (see e.g. \cite{S94a}). 
The values of \(\nu_n\) for \(n = 1,2,3,4\) are~1, 1, 6, 72.
\item
It is known that the link of the origin of phylogenetic tree space
as defined in \cite{BHV01} has the homotopy type of the
wedge of spheres. It would be interesting to work out the homotopy type of 
the link of the origin of $\mathfrak{N}(X)$, and also what other properties it might enjoy (for
example, is it Cohen-Macaulay as with the tree-space defined in \cite{BHV01}?)
\item
As was pointed out in~\cite{DDF19a}, there is a connection
between the space of circular split collections
defined in \cite{DP17a} and a certain type of unrooted
phylogenetic networks called level-1 networks.
Since these unrooted level-1 networks can
be regarded as unrooted $X$-cactuses,
it would be interesting to 
investigate if there are some connections between 
$\mathfrak{N}(X)$ and the space of circular split collections.
\item
It would be interesting to define and understand the
geometry of spaces of more complicated phylogenetic networks
with arc lengths. Two 
obvious candidates for such an investigation are rooted level-2 networks
and tree-child, time consistent networks (see \cite[Chapter 10]{S16a}
for definitions). Moreover, one could try to relax the requirement
that the phylogenetic networks are equidistant.
\item
How does the distance between equidistant \(X\)-cactuses in $\mathfrak{N}(X)$
compare to other distance measures between phylogenetic networks?  For example,
it was shown in \cite{amenta2007approximating}
that the weighted Robinson-Foulds distance between
phylogenetic trees \cite{robinson1979comparison}
is a $\sqrt{2}$-approximation of the distance between
phylogenetic trees in the tree space defined in \cite{BHV01}.
\end{itemize}

\subsubsection*{Acknowledgments}
KTH and VM thank the Department of Mathematics at
City College of New York (City University of New York)
and the American Museum of Natural History for their hospitality.
MO is partially supported by the US National Science Foundation (DMS 1847271).
This work was supported by a grant from the Simons Foundation (\#355824, Megan Owen).
KAS thanks the Simons Foundation (\#316124) and the US National Science
Foundation (\#1461094) for research and travel support.

\bibliographystyle{plain}
\bibliography{net-space}

\end{document}